\documentclass[floatfix,aps,twocolumn,floatfix,10pt,superscriptaddress,prl]{revtex4-2}

\usepackage{bm}
\usepackage{graphicx}
\usepackage{subfigure}
\usepackage{epstopdf}
\usepackage{siunitx}
\usepackage{braket}
\usepackage{tabularx}
\usepackage{blindtext}
\usepackage{amsmath}

\usepackage{color}

\def\bem#1{\begin{mathletters}\label{#1}}
\def\eml{\end{mathletters}}

\def\4#1{{\boldsymbol{#1}}}
\def\8#1{{\widetilde{#1}}}

\newcommand{\simleq}{\; \raisebox{-0.4ex}{\tiny$\stackrel
{{\textstyle<}}{\sim}$}\;}

\begin{document}

\title {Finding the nitrogen-vacancy singlet manifold energy level using charge conversion pulse sequences}

\author{I. Meirzada}
\affiliation{The Racah Institute of Physics, The Hebrew University of Jerusalem, Jerusalem 91904, Israel}

\author{S. A. Wolf}
\affiliation{The Racah Institute of Physics, The Hebrew University of Jerusalem, Jerusalem 91904, Israel}
\affiliation{The Center for Nanoscience and Nanotechnology, The Hebrew University of Jerusalem, Jerusalem 91904, Israel}

\author{N. Bar-Gill}
\affiliation{The Racah Institute of Physics, The Hebrew University of Jerusalem, Jerusalem 91904, Israel}
\affiliation{The Center for Nanoscience and Nanotechnology, The Hebrew University of Jerusalem, Jerusalem 91904, Israel}
\affiliation{Dept. of Applied Physics, Rachel and Selim School of Engineering, Hebrew University, Jerusalem 91904, Israel}

\begin{abstract}
The vast research conducted on Nitrogen-Vacancy (NV) centers in diamond in recent years opened the door for a wide range of NV based applications. However, some properties of the NV center dynamics and energy levels remain unknown. In this work, we propose a charge conversion pulsed sequence scheme for locating the NV center singlet manifold energy level, by multi-color excitation using a tunable laser. We present two approaches for readout - photo induced current and NV$^0$ population - and discuss their efficiency for different regimes in the relevant spectrum. 

\end{abstract}

\maketitle

\section{Introduction}
NV centers in diamond \cite{doherty_nitrogen-vacancy_2013} stand at the center of a variety of scientific and technological research avenues, such as quantum sensing and quantum information processing \cite{clevenson_broadband_2015,acosta_broadband_2010,dolde_electric-field_2011,taylor_high-sensitivity_2008,loretz_nanoscale_2014,trusheim_wide-field_2016}, due to their reliability and versatility of operation. These applications rely on the NV center's unique optical and spin properties, which have been studied both theoretically and experimentally \cite{Maze_2011, PhysRevB.81.041204,acosta_optical_2010,Manson_model_for_PhysRevB.74.104303,Neumann_2009,Aslam_2013}. However, some of these properties are yet to be determined. Completing this picture holds the potential for both improving the current NV based applications and discovering new opportunities using NV centers.

\begin{figure}[tbh]
{\includegraphics[width = 0.9 \linewidth]{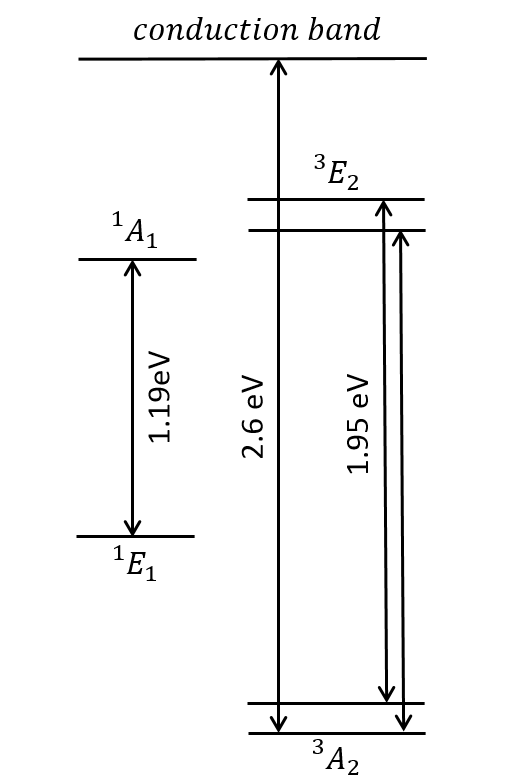}}
\caption{A simplified summary of the energy levels in the NV$^-$ and their distance from each other. The exact energy of singlet manifold ($^1A$ and $^1E$) compared to the triplet manifold ($^3E$ and $^3A$) or the conduction band is unknown.} 
\label{fig:energylevel}
\end{figure}

The NV center is composed of a nitrogen atom and a vacancy occupying two adjacent sites in the diamond lattice. The electronic ground state is a spin triplet with 2.87 GHz zero field splitting between spin projections $m_s = 0$ and $m_s = \pm 1$. The electronic excited state is composed of a highly radiative spin triplet located 1.95 eV above the triplet ground state and 0.65 eV below the conduction band, and a spin singlet with weak radiative coupling. The singlet manifold comprises two states, $^1A$ ("excited") and $^1E$ ("ground"). Although the energy gap between the two singlet states was measured to be 1.19 eV, their exact location relative to the triplet states remains unknown. Fig. \ref{fig:energylevel} summarizes the known energy gaps between the different states of the NV$^-$. 

In this work we propose a detailed scheme for identifying the singlet manifold energies. The aforementioned energies bound the energy gap between the singlet ground state and the conduction band between 1.84 eV, corresponding to 674 nm, and 2.6 eV, corresponding to 477 nm. Thus, we can divide this range into three parts, each resulting in different dynamics during the pulse sequence:
\begin{enumerate}
    
    \item 637 nm $\leq\lambda_{ion}\leq$ 674 nm - $\lambda_{ion}$ is longer than the ZPL for both NV charge states. These photons can only be absorbed by the excited states, either to be ionized (recombined) to NV$^0$ (NV$^-$) or decay back to the ground states via stimulated emission. 
    \item 575 nm $\leq\lambda_{ion}\leq$ 637 nm - $\lambda_{ion}$ is longer than the NV$^0$ ZPL and shorter then the NV$^-$ ZPL. The NV$^-$ can be excited from the ground state but not the NV$^0$. 
    \item 477 nm $ \leq  \lambda_{ion} \leq $ 575 nm - $\lambda_{ion}$ is shorter than the ZPL for both NV charge states, allowing for excitation from both ground states \cite{Aslam_2013,meirzada_negative_2017}. 
\end{enumerate}

The proposed pulse sequence aims to induce substantial ionization of the NV$^-$ ground singlet state into the neutral charge NV$^0$, thus generating additional NV$^0$ population or photo-induced current \cite{Emilie_Bourgeois_PhysRevB.95.041402}, once the threshold wavelength for ionization is reached. The timings and powers in the sequences utilize the previously published spin and charge dynamics for both charge states \cite{Robledo_2011, acosta_optical_2010, Manson_model_for_PhysRevB.74.104303, tetienne_magnetic-field-dependent_2012, chen_subdiffraction_2015, jeske_stimulated_2017,meirzada_negative_2017}. 

\begin{figure}[tbh]
{\includegraphics[width = 1\linewidth]{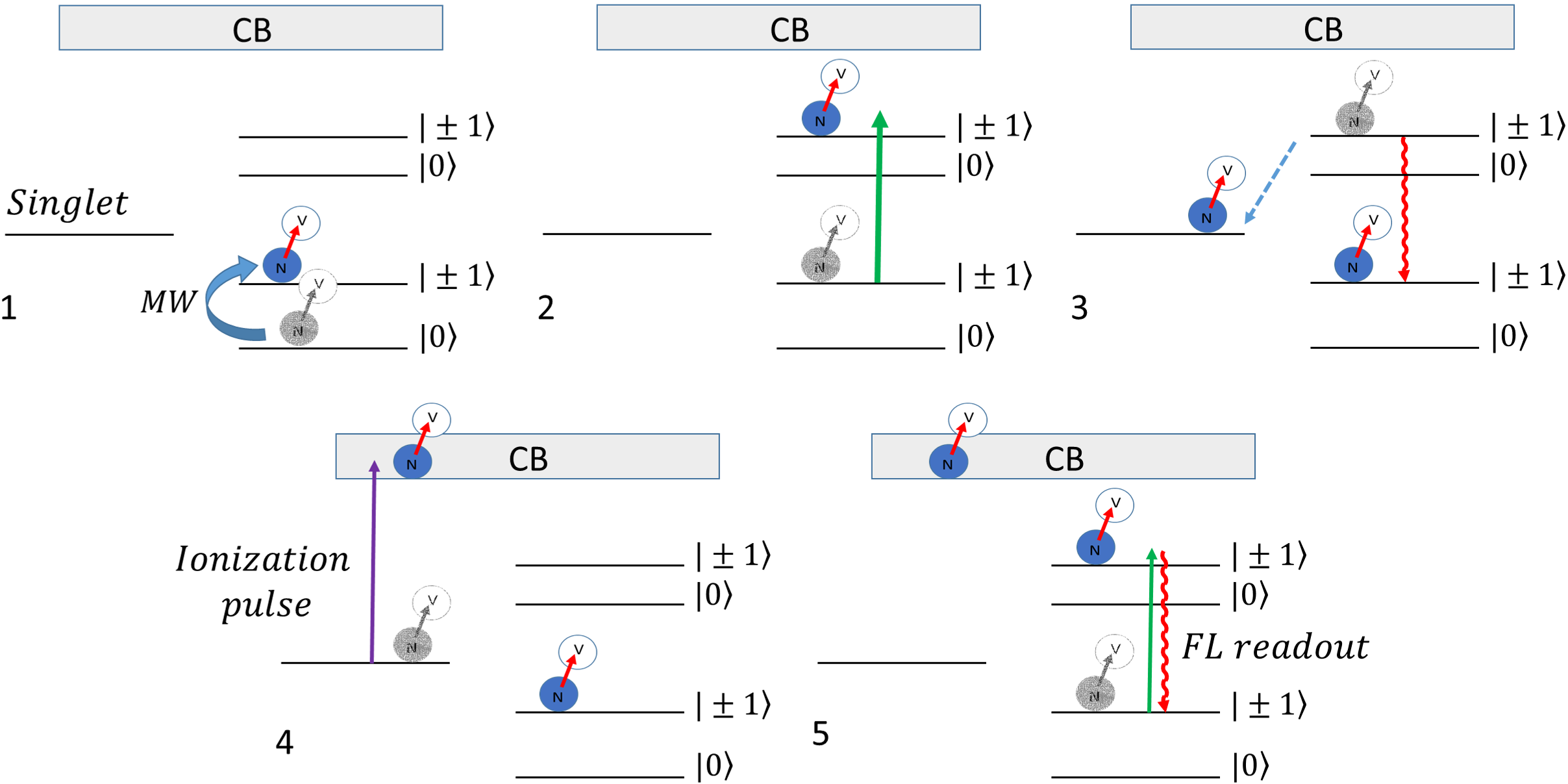}}
\caption{The seven-step sequence for identifying ionization from the singlet ground state. 
1. After initializing the NV$^-$ into the $m_s = 0$ spin state, a $\pi$ pulse is applied to initialize the NV into the $m_s = \pm$1 spin state. 
2. A short, strong green pulse populates the triplet excited state. 
3. Delay of $\sim$ 30 ns is applied for the excited state to decay to avoid ionization from the excited state. 
4. A $\sim$ 0.1 $\mu s$, second pulse (ionization pulse) is applied while collecting photo-current. 
5. (For fluorescence based measurement) green laser excitation while collecting fluorescence. } 
\label{fig:pulse sequence}
\end{figure}

The pulse sequence, described in Fig. \ref{fig:pulse sequence}, contains six steps. We start with initialization of the NV$^-$ to the $m_s = 0$ state, followed by a microwave (MW) $\pi$-pulse to initialize the NV into the $m_s = \pm 1$ state. Then, a short ($\sim$ 30 ns) green excitation populates the triplet excited state ("population pulse"). Preparing the NV in the spin 1 state will help populate the singlet state, thus enhancing the sensitivity to ionization from the singlet. A short delay allows the NV to either decay to the singlet level or back to the triplet ground state, thus preventing unwanted ionization from the triplet excited state while having a negligible effect on the singlet ground state population due to its metastable nature. Then a stronger ionization pulse (477 nm $\leq\lambda\leq$ 674 nm) is applied while photo-current is collected from the NV. A fluorescence based measurement requires an additional readout pulse for collecting photons. Unless specified otherwise, the population and ionization pulses powers are 2 mW and 5 mW, respectively. The readout pulse power is 100 $\mu$W, with a 1$\mu$s duration. 

In the results given below, each simulation presents two curves. A red curve, in which the ionization laser energy is higher than the singlet ionization threshold ($ \lambda_{ion} \leq \lambda_s$) and a blue curve, in which the ionization laser energy is lower than the singlet ionization threshold ($\lambda_s \leq  \lambda_{ion}$). In cases for which 532 nm $\leq  \lambda_{ion}$ the simulations take into account that the pulses used for initialization and population ionize the singlet, for both the red and blue curves. 

\subsection {Fluorescence-based measurements}

In the fluorescence-based measurements, reaching the ionization threshold manifests in the final NV$^-$ population, which can be accurately measured \cite{Lukin_SCC_PhysRevLett.114.136402,hopper_amplified_2018}. 
We start with the longer wavelengths, for which the threshold ionization does not excite the triplet ground state. In this case the green excitation laser (532 nm) also ionizes the NV from the singlet ground state. The following figures draw the expected NV$^-$ population under different conditions such as laser power, ionization duration, delay time and ionization cross section, outlining the parameter space in which the pulse sequence results in strong evidence that the wavelength threshold $\lambda_s$ for singlet ionization is reached. All laser power parameters in this work assume diffraction limited excitation beams. 

\begin{figure}[tbh]
\subfigure[]
{\includegraphics[width = 0.45 \linewidth]{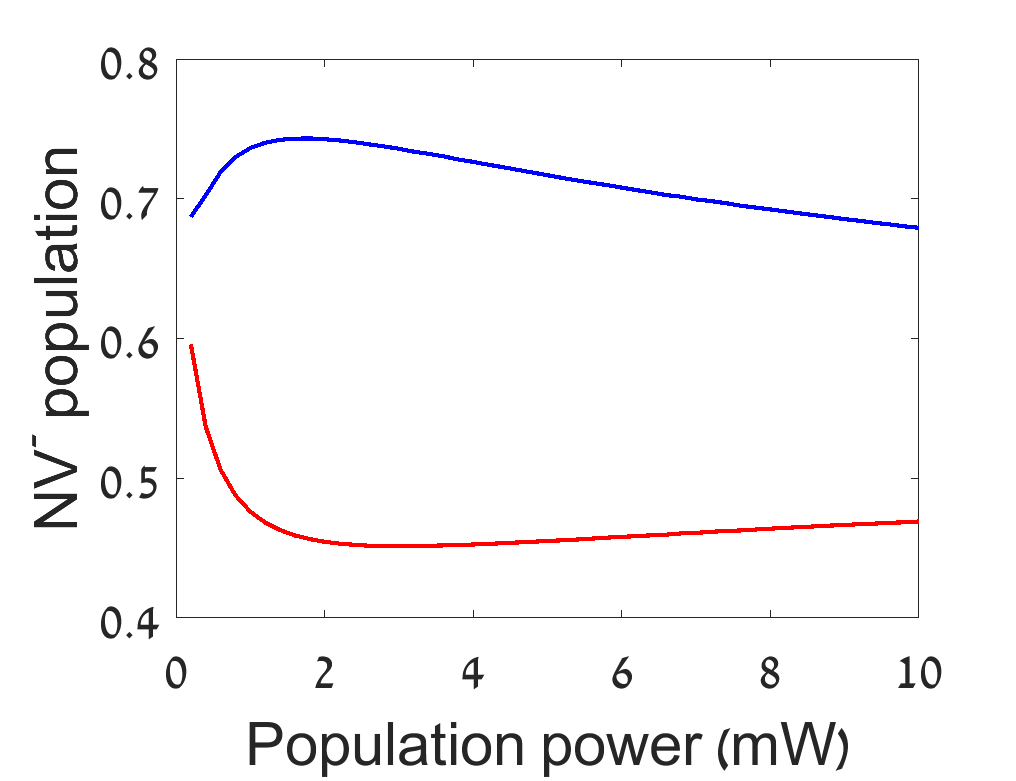}}
\subfigure[]
{\includegraphics[width = 0.45 \linewidth]{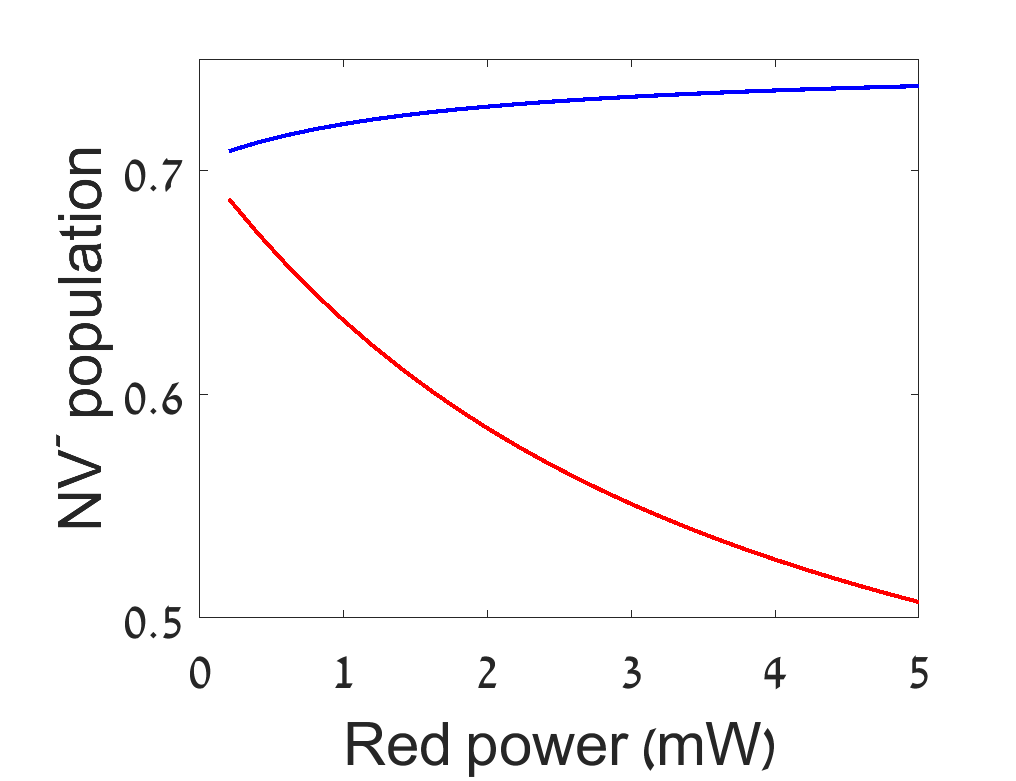}}
\subfigure[]
{\includegraphics[width = 0.45 \linewidth]{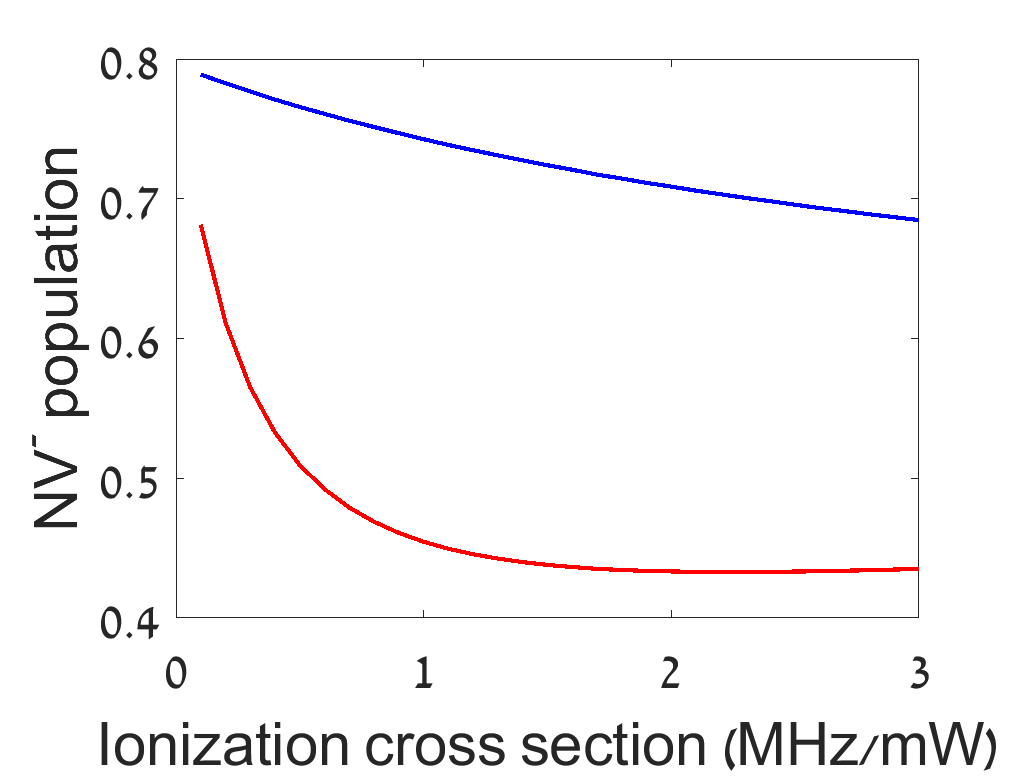}}
\subfigure[]
{\includegraphics[width = 0.45 \linewidth]{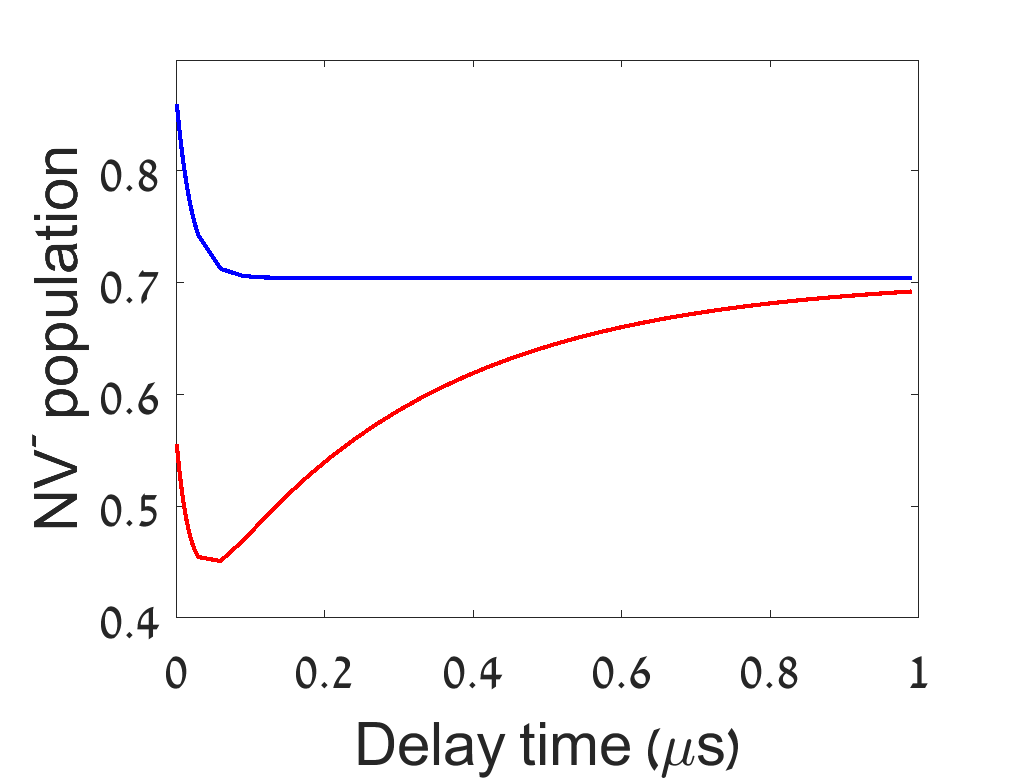}}
\subfigure[]
{\includegraphics[width = 0.45 \linewidth]{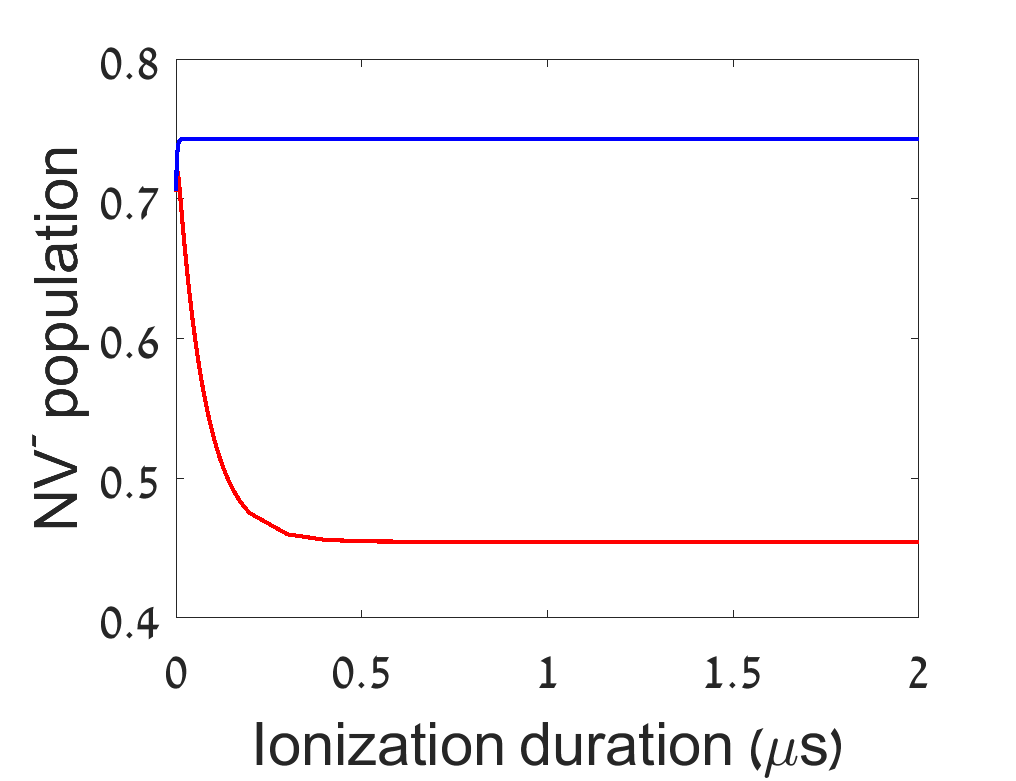}}

\caption{NV$^-$ final population for various sequence parameters where 637 nm $ \leq  \lambda_{ion} \leq $ 674 nm. a. NV$^-$ population vs green power, populating the singlet level. A power $\leq$ 1 mW is preferred for significant contrast. b. NV$^-$ population vs red excitation power. Higher power provides higher contrast, due to the delay between the green population pulse and the ionization pulse. c. NV$^-$ population vs singlet ionization cross section (in MHz/mW). Ionization rate higher than 0.5 MHz/mW is expected to result in detectable contrast. d. NV$^-$ population vs the delay between the population and ionization pulses. For short times, the recombination from NV$^0$ builds higher population in the negative state. After a few tens of ns the excited states decay, and the contrast between the two cases decreases with the singlet state lifetime. e. NV$^-$ population vs ionization pulse duration. The contrast increases as long as the singlet state is occupied, and doesn't decrease afterwards since no other ionization mechanism exists.}
\label{fig:PL_Red}
\end{figure}

The first set of simulations addresses the case in which the ionization threshold wavelength does not have enough energy to excite the NV$^-$ triplet ground state ("Red" range). Fig. \ref{fig:PL_Red}(a) presents the NV$^-$ population as a function of green population power. Weak powers create a small population in the singlet, while very strong powers ionize the singlet, thus also resulting in small populations in this state. Fig. \ref{fig:PL_Red}(b) presents the NV$^-$ population vs. red ionization power. The red ionization power reduces the population significantly as power increases (red line), although mild recombination (NV$^0 \rightarrow $ NV$^-$) occurs at the same time from leftover populations in the NV$^0$ excited state, as illustrated by the blue line. Fig. \ref{fig:PL_Red}(c) presents the NV$^-$ population as a function of singlet ionization cross section (expressed in units of MHz/mW) for a constant red ionization power of 10 mW, assuming the same cross section for both green and red excitation. Here the red curve decreases faster, since the green-induced ionization changes in addition to the red-induced ionization, unlike Fig. \ref{fig:PL_Red}(b) for which only the red-induced ionization changes with the increase in laser power. The blue curve decreases for the same reason. The excitation in the range 637 nm $ \leq \lambda \leq $ 674 nm can also deplete electrons from the triplet excited state directly to the triplet ground state, reducing the population needed in the singlet state for the ionization detection. Fig. \ref{fig:PL_Red}(d) shows the population dependence on the delay between the initialization and ionization pulses, providing strong evidence that the electron is ionized from the singlet state, since it's timescale changes according to the singlet-state lifetime assumed in the simulation. The small decreasing trend found for short delay times stems from ionizing the electron from the excited state, before it decays to the singlet and ground triplet states. Lastly, Fig. \ref{fig:PL_Red}(e) depicts the NV$^-$ behavior as a function of ionization duration. The timescale here is a function of both power and the singlet state lifetime. The weaker the ionization would be (due to red power or ionization cross section), the closer the timescale would be to the singlet state lifetime. 

\begin{figure}[tbh]
\subfigure[]
{\includegraphics[width = 0.45 \linewidth]{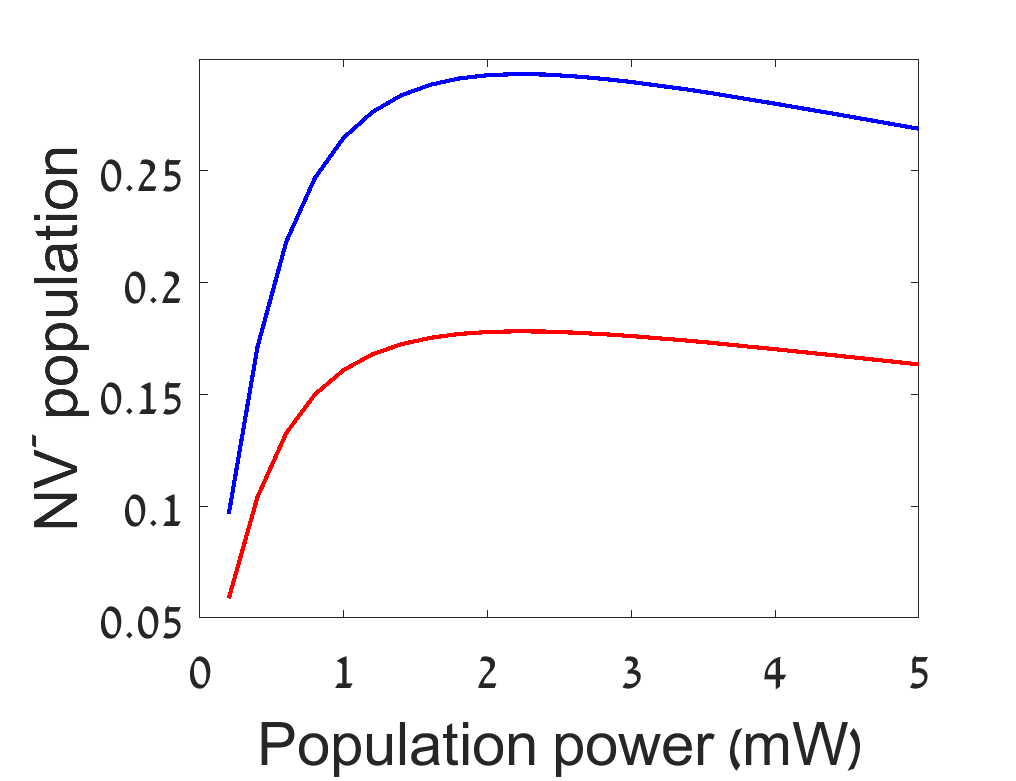}}
\subfigure[]
{\includegraphics[width = 0.45 \linewidth]{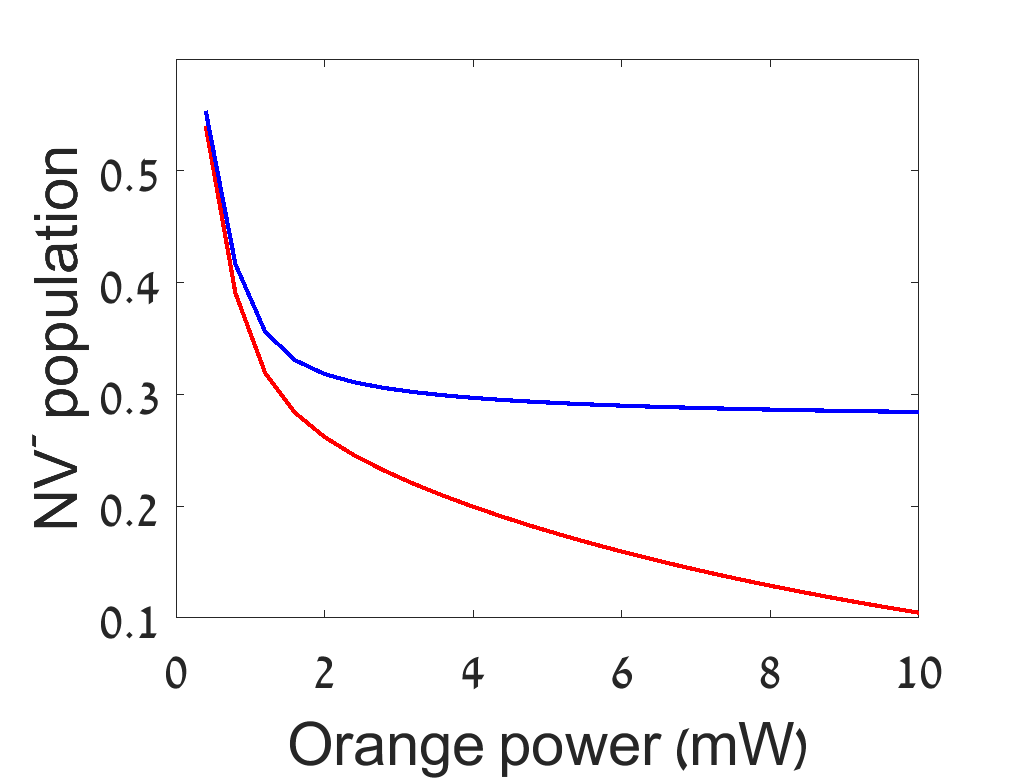}}
\subfigure[]
{\includegraphics[width = 0.45 \linewidth]{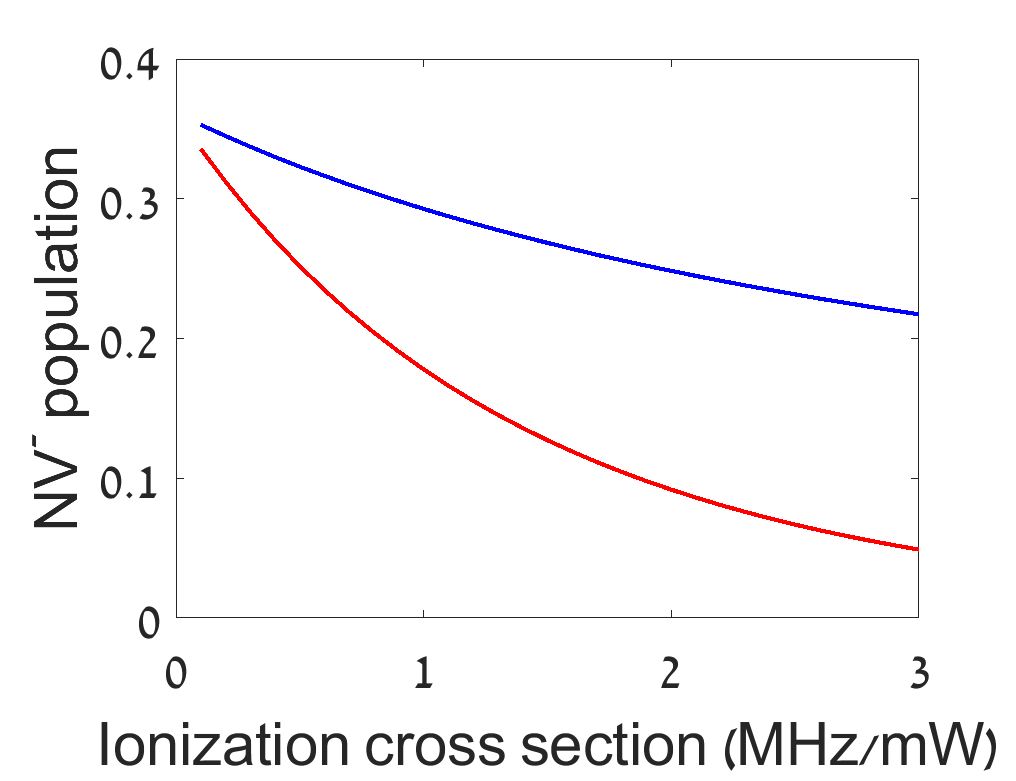}}
\subfigure[]
{\includegraphics[width = 0.45 \linewidth]{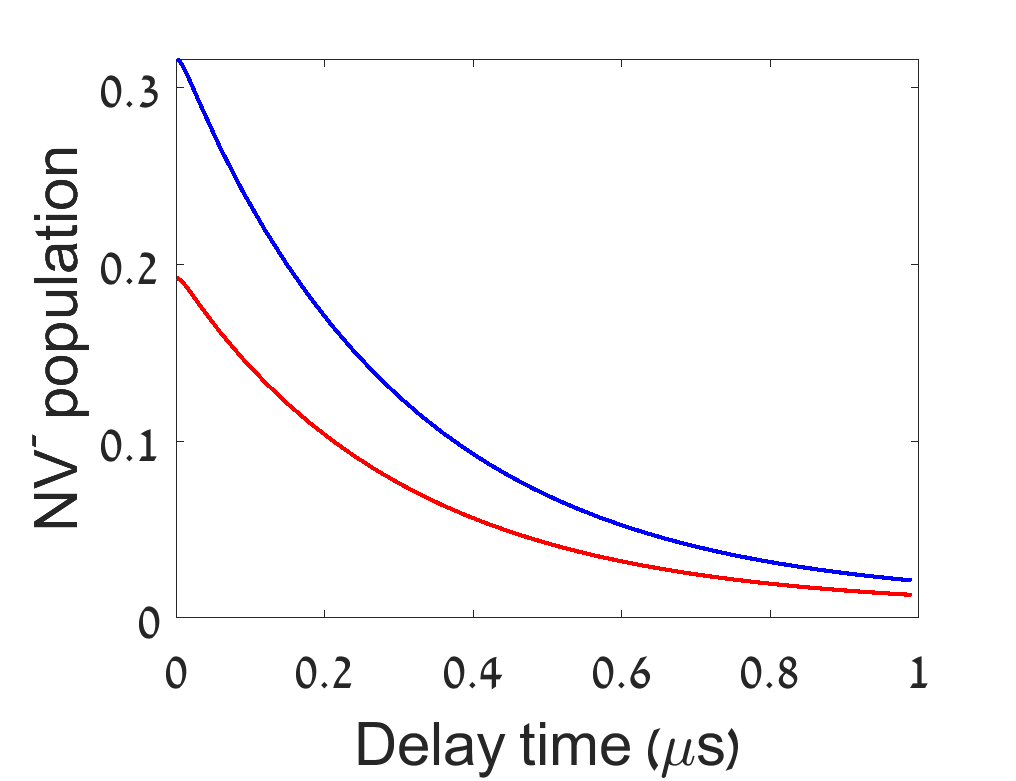}}
\subfigure[]
{\includegraphics[width = 0.45 \linewidth]{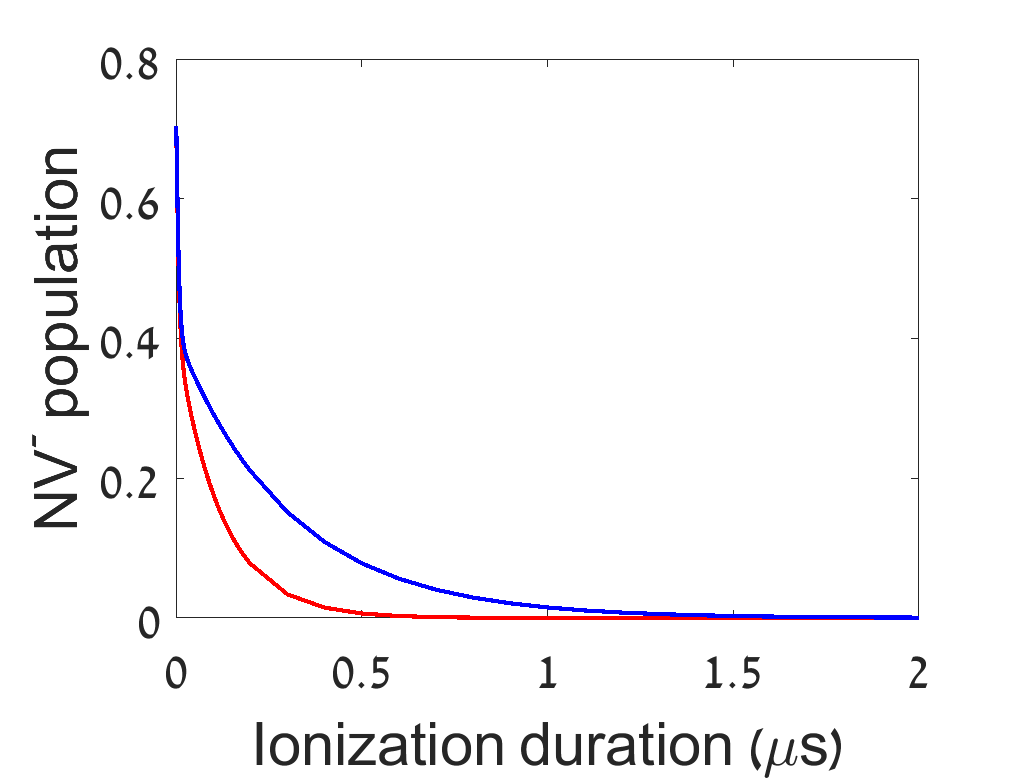}}

\caption{NV$^-$ final population for various sequence parameters, where 637 nm $\leq \lambda_s \leq $ 575 nm. a. NV$^-$ population vs green population power. A power $\leq$ 1 mW is preferred for significant contrast. b. NV$^-$ population vs orange ionization power. Higher power provides higher contrast, as the population in the triplet is expected to empty before the population in the singlet, due to the assumption of low ionization cross section of the singlet. c. NV$^-$ population vs singlet ionization cross section (in MHz/mW). Ionization cross section higher than 0.5 MHz/mW is expected to result in detectable contrast. d. NV$^-$ population vs the delay between the population and ionization pulses. The contrast decreases with a time scale of the singlet lifetime e. NV$^-$ population vs ionization pulse duration. The ionization duration must be similar to the singlet lifetime, otherwise no significant contrast is expected.} 

\label{fig:PL_Orange}
\end{figure}

The next set of figures analyzes the case in which the threshold wavelength has enough energy to excite the NV$^-$ triplet ground state, although insufficient to excite the NV$^0$ ground state ("Orange" range). In this case, the green excitation laser still ionizes the singlet. We note that in this range the steady-state population while using the ionization laser strongly favors the neutral charge state. Fig. \ref{fig:PL_Orange}(a) depicts the effect of the green excitation power on the final NV$^-$ population, suggesting that it will be good to work around 2 mW. Changing the ionization laser power creates a more complex picture, as shown in Fig. \ref{fig:PL_Orange}(b). The curves seem independent of ionization at low powers, while higher laser powers split the curves significantly. This is because the ionization cross section from the ground state is much higher than that from the singlet. At high powers the blue curve reaches a plateau, since the population in the ground state has been completely ionized. At this regime the singlet ionization becomes stronger, resulting in a splitting between the red and blue curves. The behavior of the population vs. singlet ionization cross section [Fig. \ref{fig:PL_Orange}(c)] is pretty straight forward. An ionization cross section equivalent to a rate of 1 MHz/mW should suffice for detecting the singlet ionization. The fact that the ionization laser excites the NV$^-$ manifests in Fig. \ref{fig:PL_Orange}(d), as the curves show different behaviour compared to Fig. \ref{fig:PL_Red}(d). 
Both curves change with the timescale of the singlet lifetime, yet with different coefficients, due to the additional ionization process from the singlet state. 
The unique nature of the orange wavelength for the NV system further expresses itself in Fig. \ref{fig:PL_Orange}(e), as it leaves us a very limited time interval for ionization, before most of the NV$^-$ population diminishes due to two-step ionization from the ground state. However, this behavior results in high contrast in the population difference between the two scenarios, thus making it easier to distinguish whether ionization from the singlet happens given enough signal from the NVs. Naturally, this time window depends strongly on the excitation power. In addition, both curves present a double exponential behavior: A standard, fast dynamic originated in the two-step ionization of the triplet (for both figures), followed by a distinct, slower process, expressing the effective singlet lifetime during the ionization process. The red curve decreases faster than the blue curve, suggesting that an additional process clears the population in the singlet manifold, as expected. 


\begin{figure}[tbh]
\subfigure[]
{\includegraphics[width = 0.45 \linewidth]{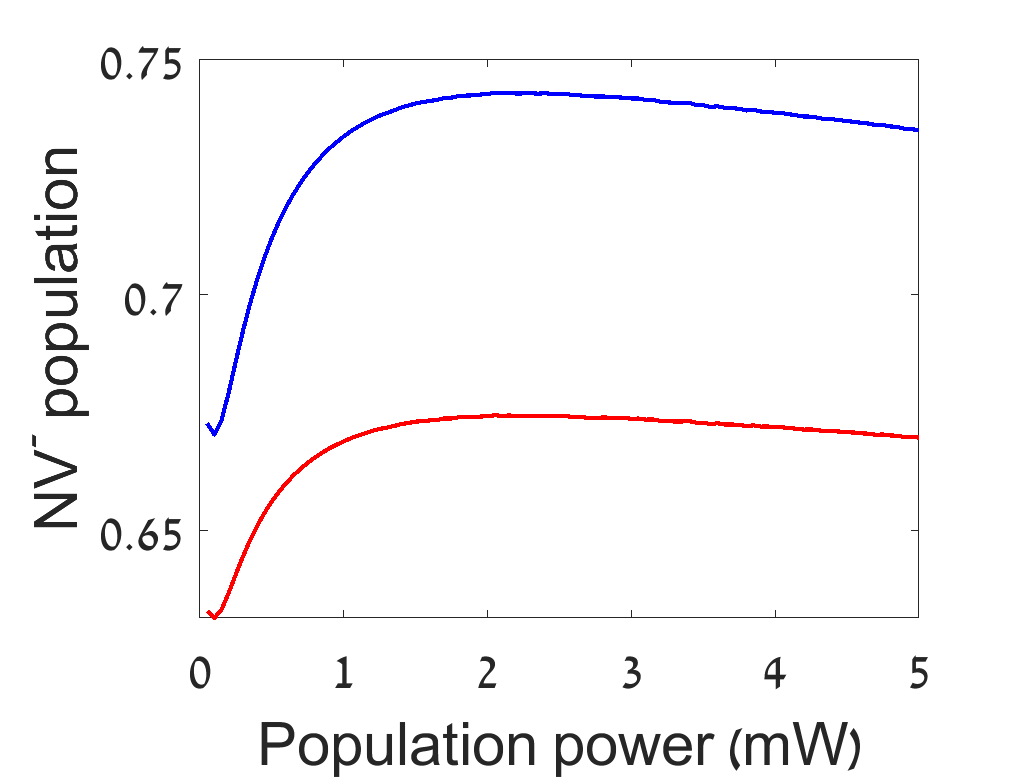}}
\subfigure[]
{\includegraphics[width = 0.45 \linewidth]{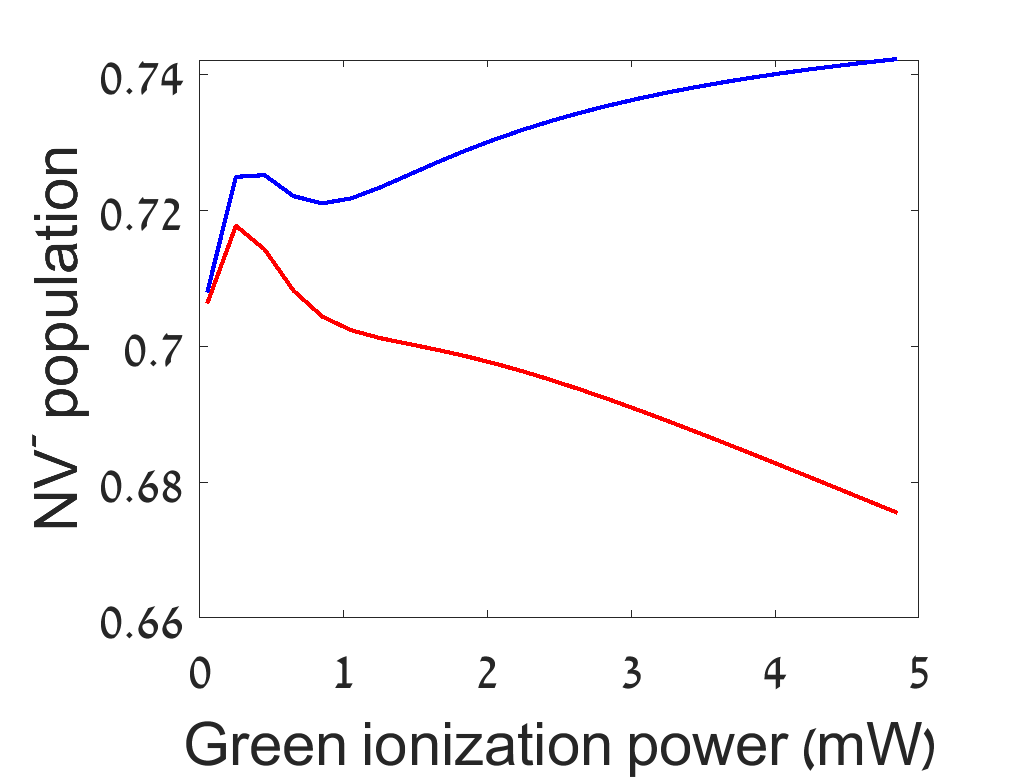}}
\subfigure[]
{\includegraphics[width = 0.45 \linewidth]{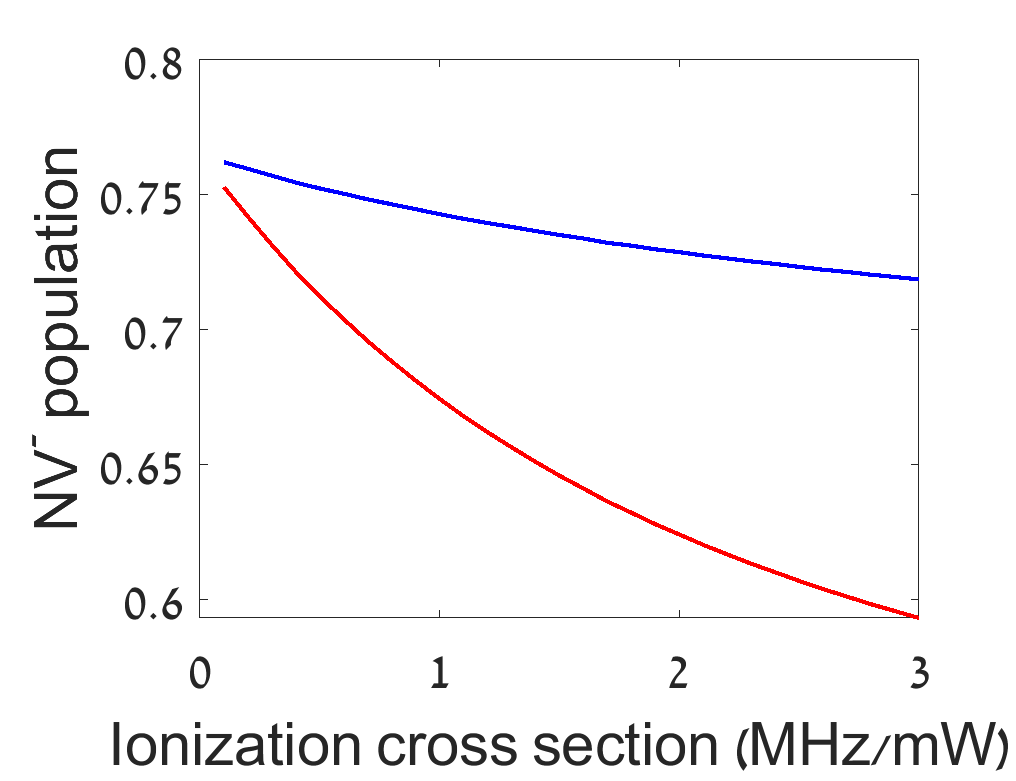}}
\subfigure[]
{\includegraphics[width = 0.45 \linewidth]{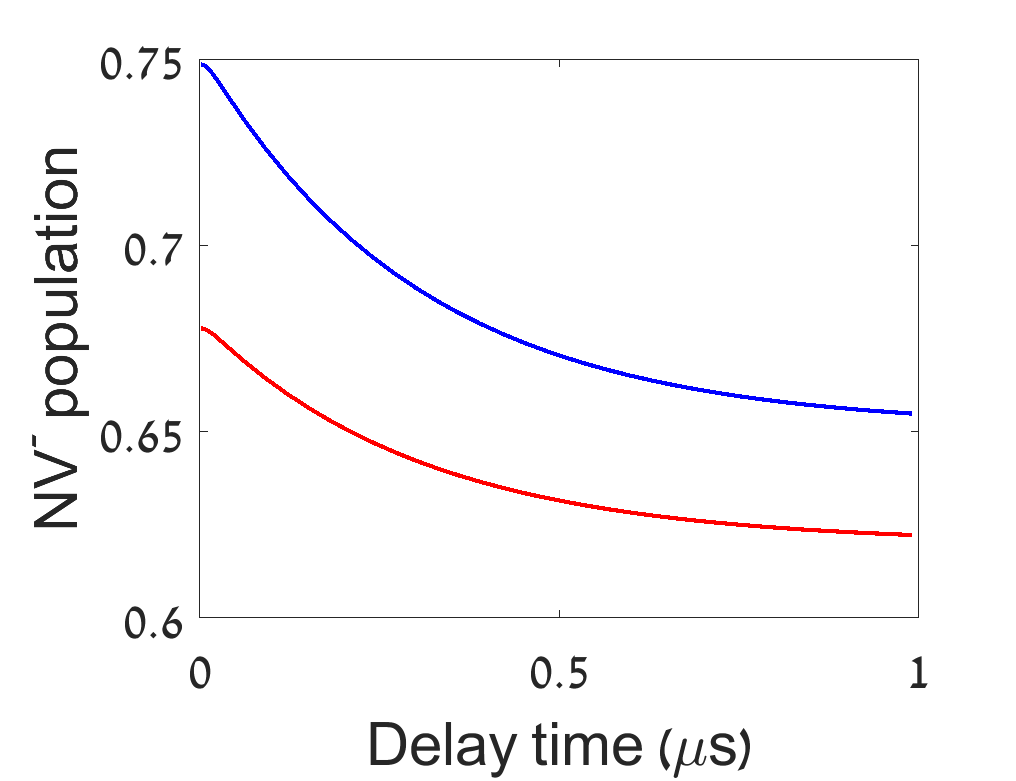}}
\subfigure[]
{\includegraphics[width = 0.45 \linewidth]{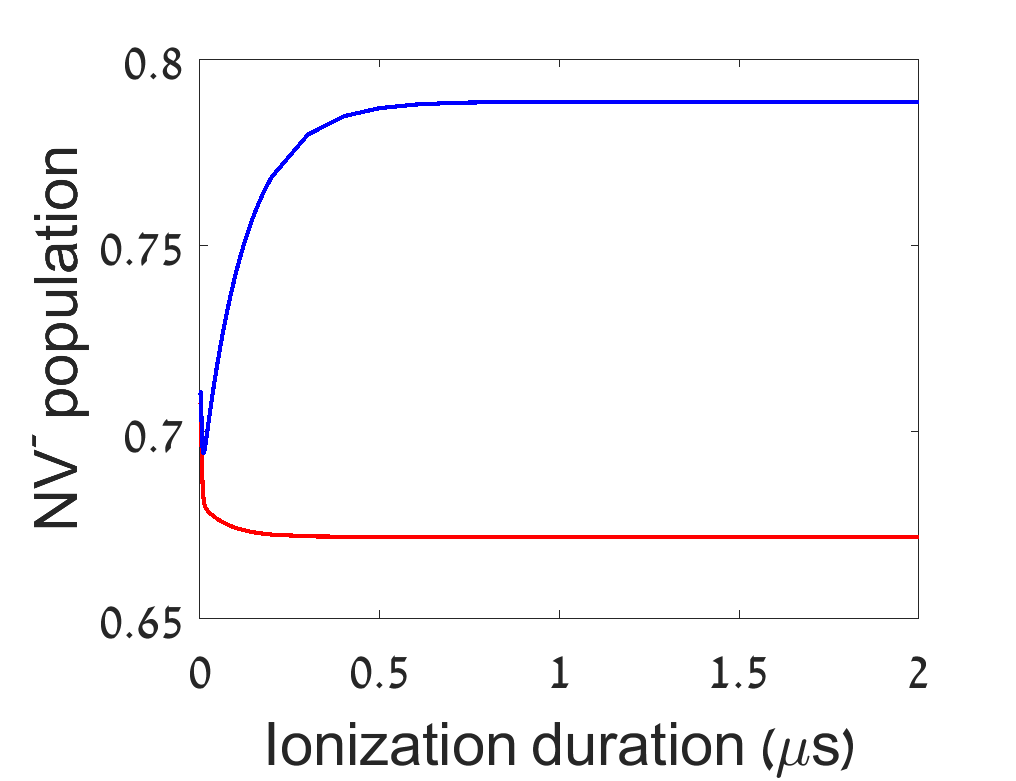}}

\caption{NV$^-$ final population for various sequence parameters when 532 nm $ \leq \lambda_s \leq $ 575 nm. a. NV$^-$ population vs green population pulse power. A power $\leq$ 1 mW is preferred for significant contrast. b. NV$^-$ population vs the green ionization power. Higher power provides higher contrast, as the singlet ionization cross section is expected to be smaller that the excited state cross section. c. NV$^-$ population vs singlet ionization cross section (in MHz/mW). Ionization rate higher than 0.5 MHz/mW is expected to result in detectable contrast. d. NV$^-$ population vs the delay between the population and ionization pulses. The contrast decreases with a time scale of the singlet lifetime e. NV$^-$ population vs ionization pulse duration. The ionization duration is not bounded from above due to the fact that the populations differences are not transient.} 
\label{fig:PL_Green}
\end{figure}

Moving forward to the next wavelength range, 575 nm $\leq \lambda_s \leq $ 532 nm ("Green" range), in which the ionization laser excites both ground states. The steady-state population, in this case, favors the negative charge state even during the ionization pulse. We note that the lower bound of this range is dictated by the initialization and population wavelength and not an intrinsic property of the NV center.
Fig. \ref{fig:PL_Green}(a) and \ref{fig:PL_Green}(b) draw the NV$^-$ population as a function of population and ionization pulse powers, respectively. Both suggest using power $\geq$ 2 mW for a substantial decrease in the final NV$^-$ population if singlet ionization occurs. This stems from the requirement for substantial population in the singlet manifold, and due to the assumption that the singlet ionization cross section is much smaller than the excited state cross section. The increase in NV$^-$ population in Fig. \ref{fig:PL_Green}(a) comes from the way the sequence is designed. The initialization pulse creates substantial population in the singlet state. During the ionization pulse ($< 300$ ns), this population does not play a role in the charge dynamics, resulting in a new steady state between the charge states, favoring NV$^-$. At high ionization laser powers, however, the rate from the NV$^-$ excited state to the singlet becomes negligible compared to the ionization rate and causes a decrease in NV$^-$ population. This same mechanism also explains the behavior of the blue curve. For low powers, the decay from the singlet back to the triplet state is faster than the charge dynamics between NV$^-$ and NV$^0$. Using powers around 1 mW, the ionization rate from the triplet exceeds the decay to the singlet state, while not being strong enough to take advantage of the population already in the singlet state.
Similarly to Fig. \ref{fig:PL_Orange}(c), an ionization cross section higher than 0.5 MHz/mW should suffice for detecting ionization, as depicted in Fig. \ref{fig:PL_Green}(c). The NV$^-$ population as a function of delay time, shown in Fig. \ref{fig:PL_Green}(d), behaves similarly to Fig. \ref{fig:PL_Orange}(d), with a larger NV$^-$ population. Fig. \ref{fig:PL_Green}(e) illustrates the qualitative difference of using a wavelength that excites both charge states. The ionization pulse duration no longer has an upper limit, as the contrast in the populations exists as a difference between the two steady-state populations, as opposed to Fig. \ref{fig:PL_Orange}(e) in which the difference in populations is transient. The blue curve increases, since the previous, population pulse, also ionized the singlet, thus the steady-state population increases during this step. For the same reason, the NV$^-$ population does not display a significant difference in the red curve. The fast dynamics stem from the delay between the population and ionization pulses. 

\begin{figure}[tbh]
\subfigure[]
{\includegraphics[width = 0.45 \linewidth]{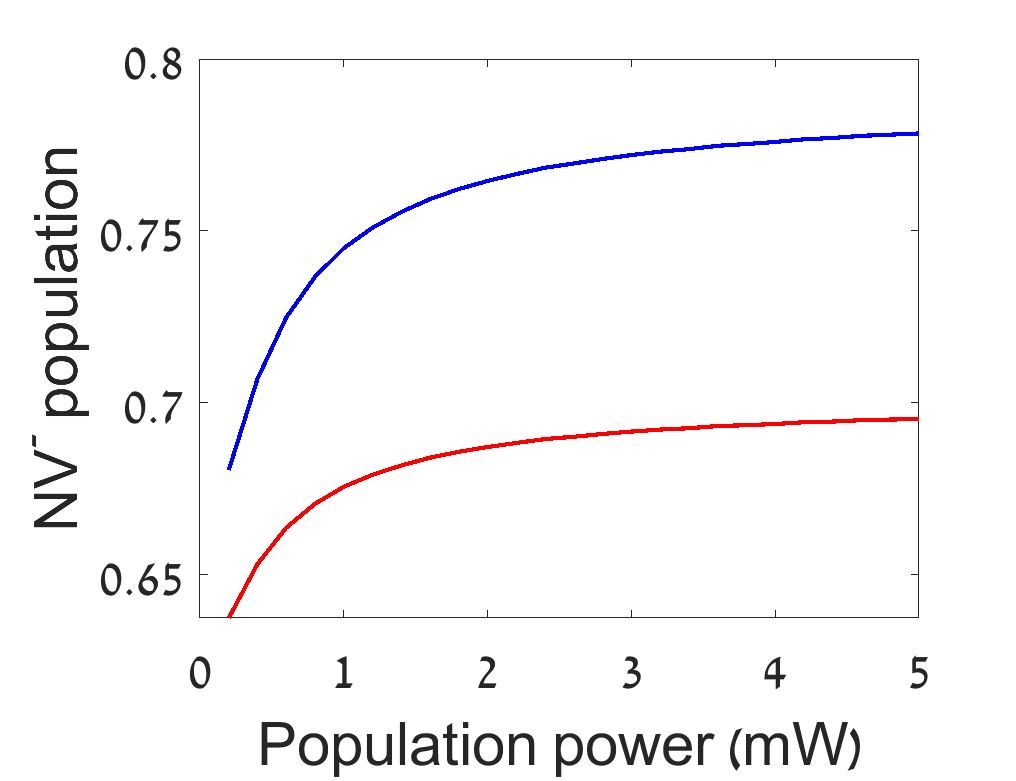}}
\subfigure[]
{\includegraphics[width = 0.45 \linewidth]{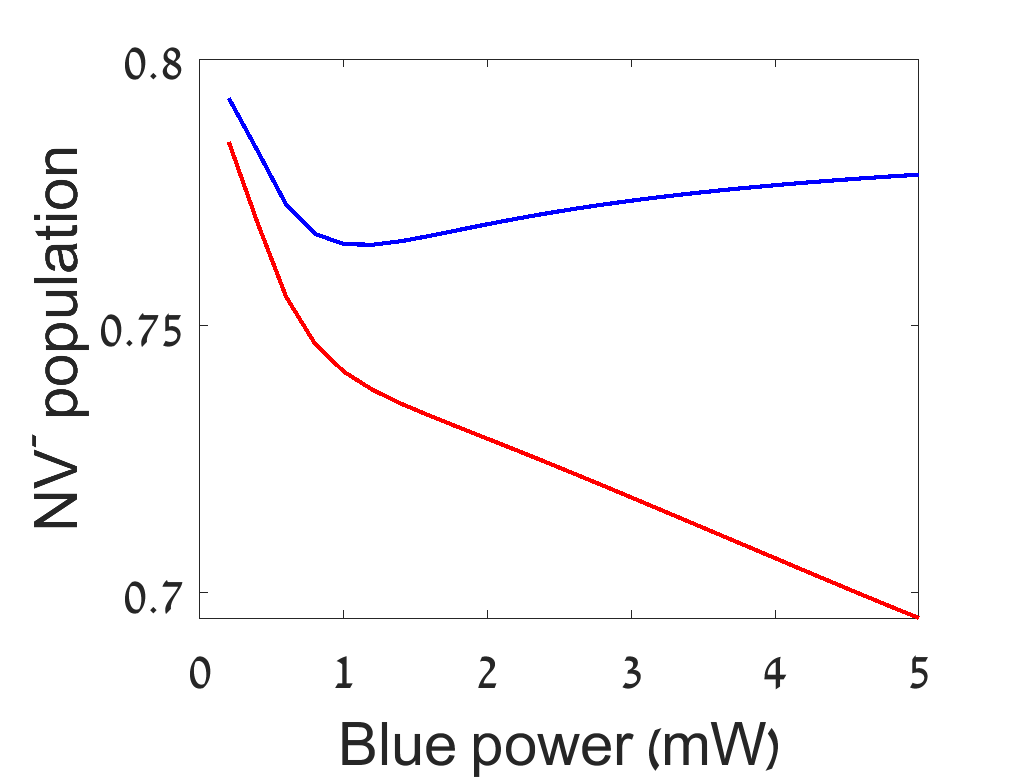}}
\subfigure[]
{\includegraphics[width = 0.45 \linewidth]{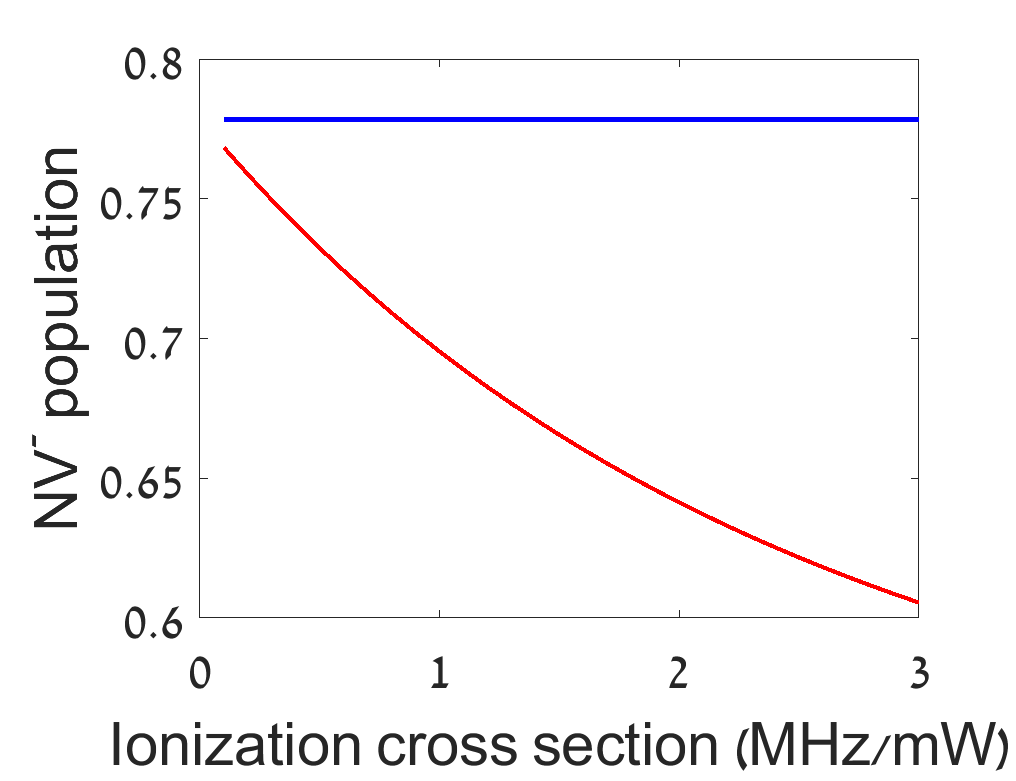}}
\subfigure[]
{\includegraphics[width = 0.45 \linewidth]{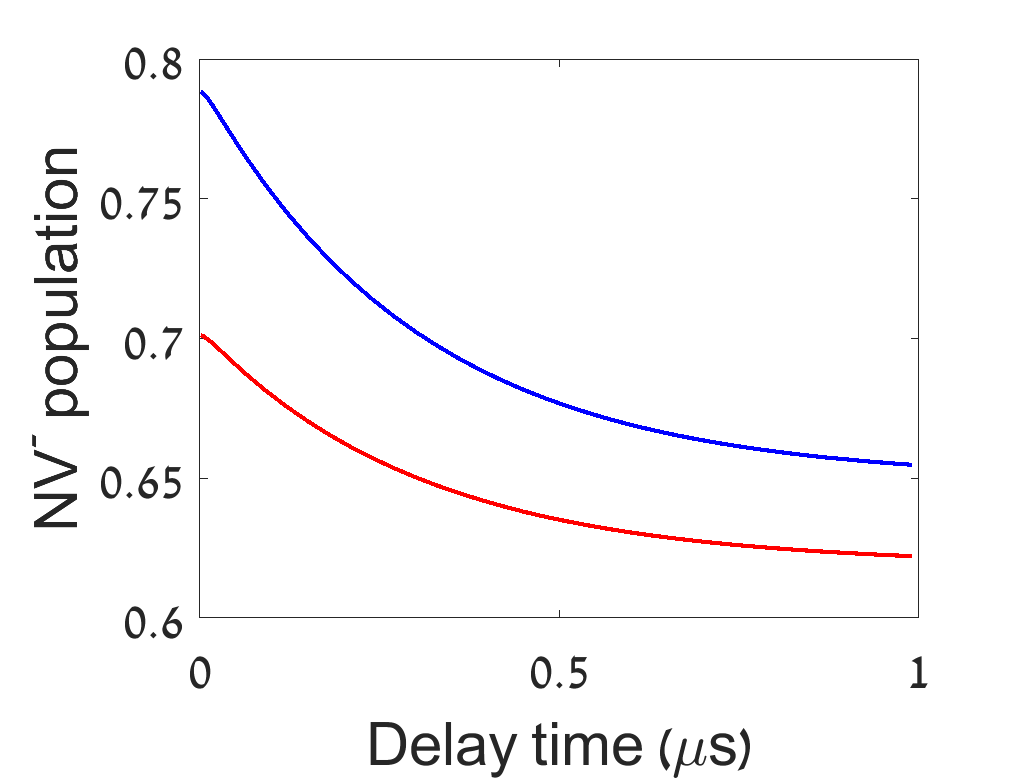}}
\subfigure[]
{\includegraphics[width = 0.45 \linewidth]{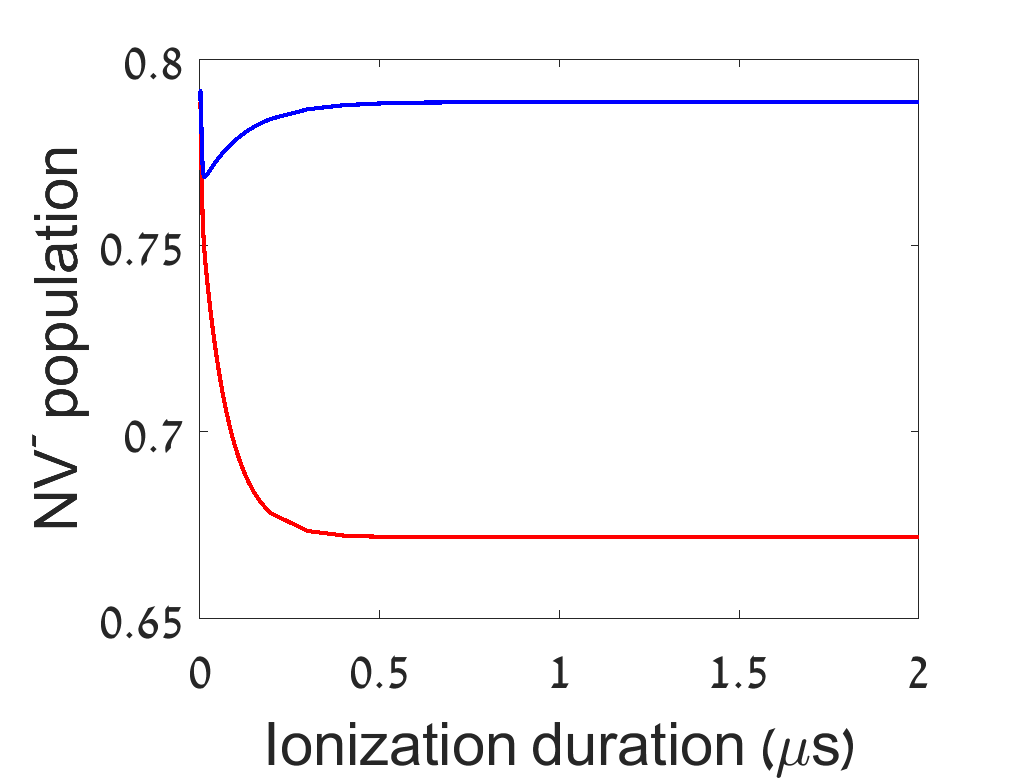}}

\caption{
NV$^-$ final population for various sequence parameters when 477 nm $ \leq \lambda_s \leq $ 532 nm, showing almost identical behavior as in Fig. \ref{fig:PL_Green}. a. NV$^-$ population vs green population pulse power. A power $\leq$ 1 mW is preferred for significant contrast. b. NV$^-$ population vs the green ionization power. Higher power provides higher contrast, as the singlet ionization cross section is expected to be smaller than the excited state cross section. c. NV$^-$ population vs singlet ionization cross section (in MHz/mW). Ionization rate higher than 0.5 MHz/mW is expected to result in detectable contrast. d. NV$^-$ population vs the delay between the population and ionization pulses. The fast dynamics (short delays) stem from ionization from the triplet states. The contrast decreases with a time scale of the singlet lifetime. e. NV$^-$ population vs ionization pulse duration. The ionization duration is not bounded from above due to the fact that the populations differences are not transient.} 

\label{fig:PL_Blue}
\end{figure}

Up to now, the population laser also ionized the NV$^-$ singlet state, as it was more energetic than the threshold wavelength for ionization. The last wavelength range, 532 nm $\leq \lambda_s \leq $ 477 nm ("Blue" range), presents a different situation, in which 
the population laser does not ionize the NV$^-$ singlet state while still exciting both ground states. For this reason, the population power simulated in this set of simulations was 5 mW (instead of 2 mW used in Figs. \ref{fig:PC_Red}-\ref{fig:PC_Green}), in order to reach higher population in the singlet. Fig. \ref{fig:PL_Blue}(a) shows the typical behavior as in the previous figures. Fig \ref{fig:PC_Blue}(b) behaves like a combination of Figs. \ref{fig:PC_Orange}(b) and \ref{fig:PC_Green}(b). For lower powers the singlet becomes less populated, causing a preference for the neutral state compared to the higher power used in the population step. Fig \ref{fig:PL_Blue}(c) shows the expected behavior for cases in which the population pulse does not ionize the singlet, and Figs. \ref{fig:PC_Blue}(d) and \ref{fig:PC_Blue}(e) resemble \ref{fig:PC_Green}(d) and \ref{fig:PC_Green}(e) with slight changes for the same reason. 

\subsection{photo-current based measurement}

Measuring using photo-current (PC) provides a 'direct' look into the charge dynamics of the system, as it isolates the signal coming directly from the NV$^- \rightarrow$ NV$^0$ process, instead of relying on transient or steady-state populations in one of the charge states. As opposed to fluorescence-based measurements, in this case no-ionization results in no signal in the system (up to background noise). On the other hand, as the energy increases, other defects may cause a non-negligible background current which can mask the signal from the singlet \cite{Emilie_Bourgeois_PhysRevB.95.041402}. The PC signal is calculated as the (ionization rate) x (population) in the relevant states (triplet excited state and singlet ground state), and is given in arbitrary units (AU). 

\begin{figure}[tbh]
\subfigure[]
{\includegraphics[width = 0.45 \linewidth]{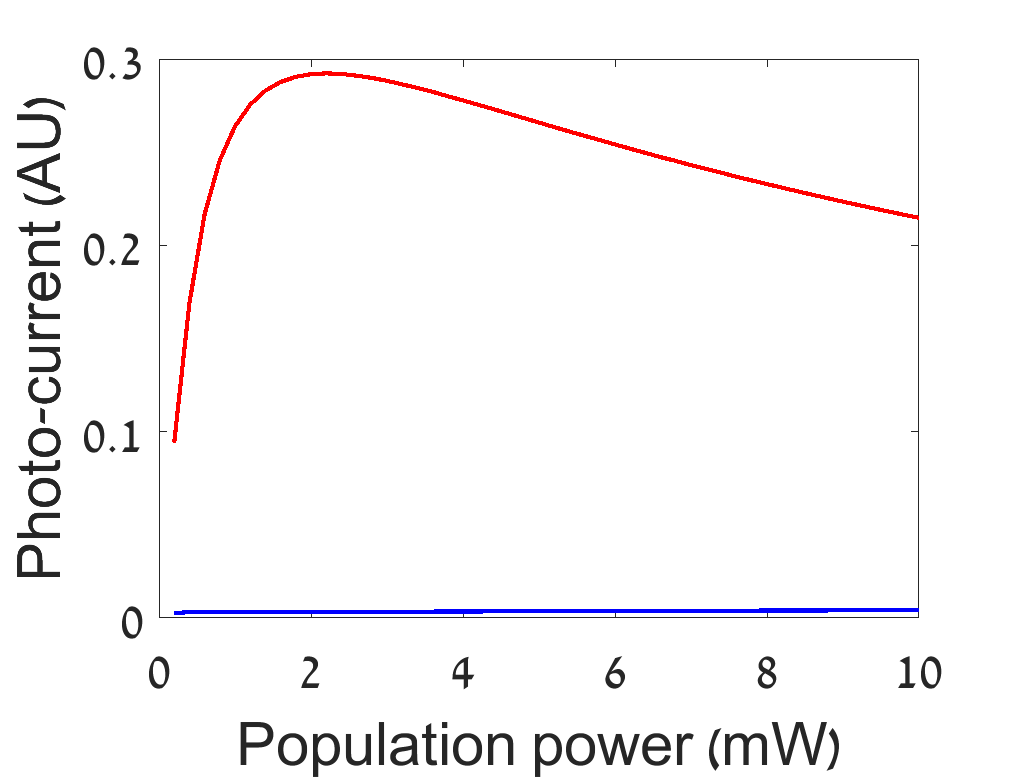}}
\subfigure[]
{\includegraphics[width = 0.45 \linewidth]{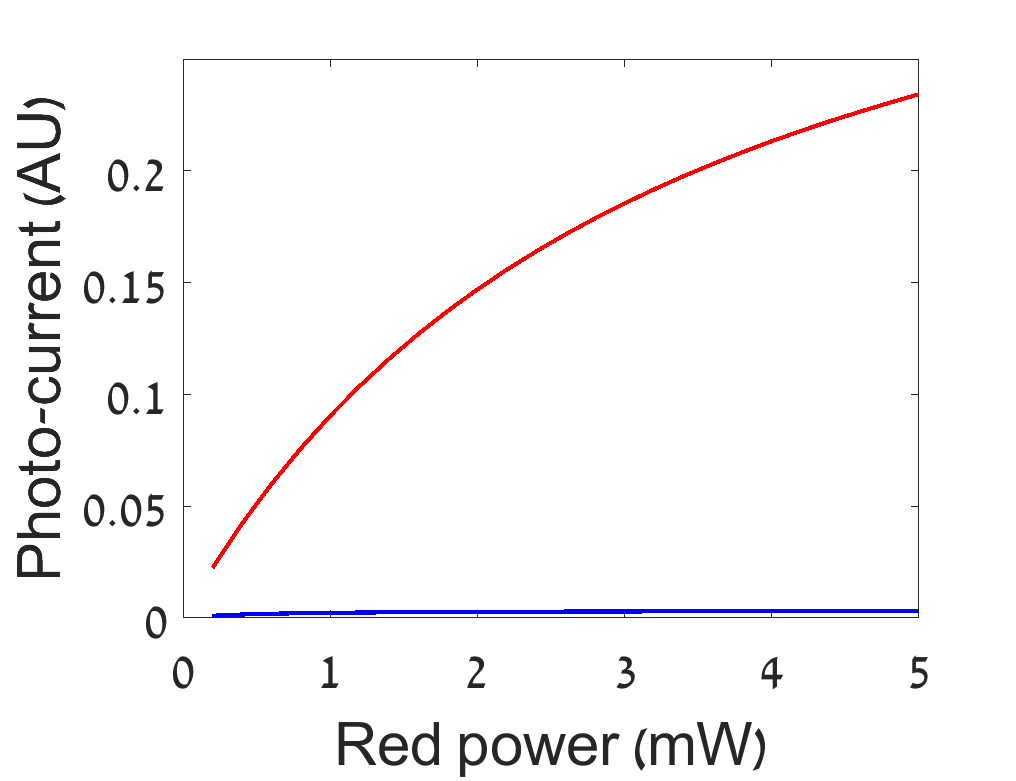}}
\subfigure[]
{\includegraphics[width = 0.45 \linewidth]{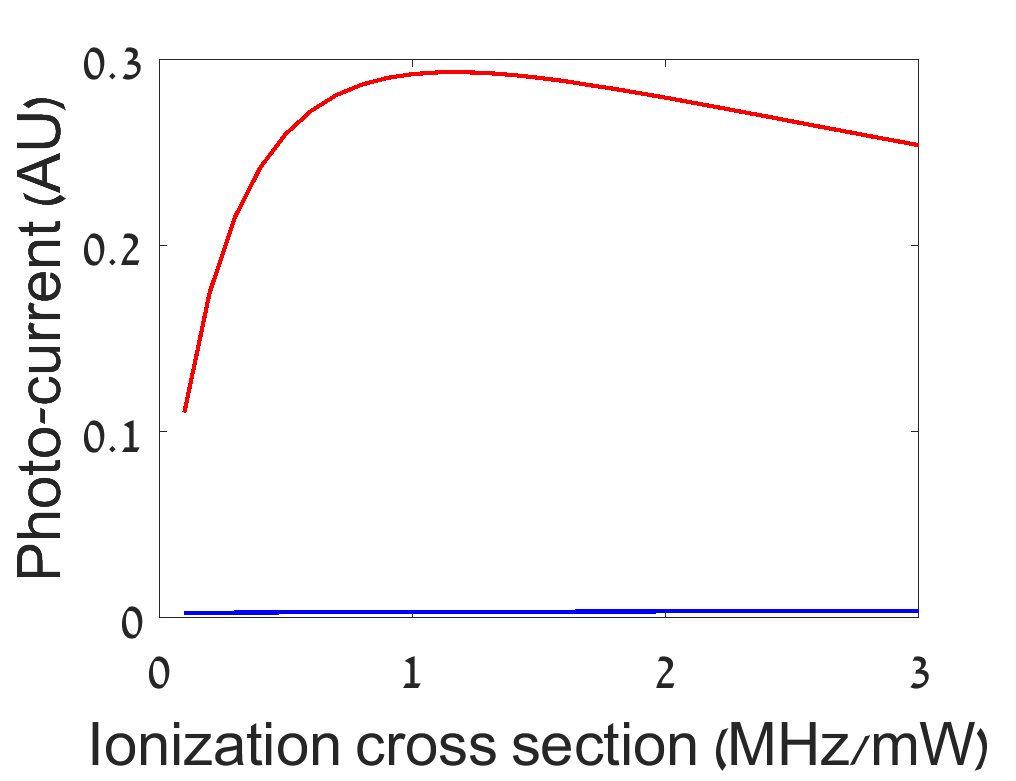}}
\subfigure[]
{\includegraphics[width = 0.45 \linewidth]{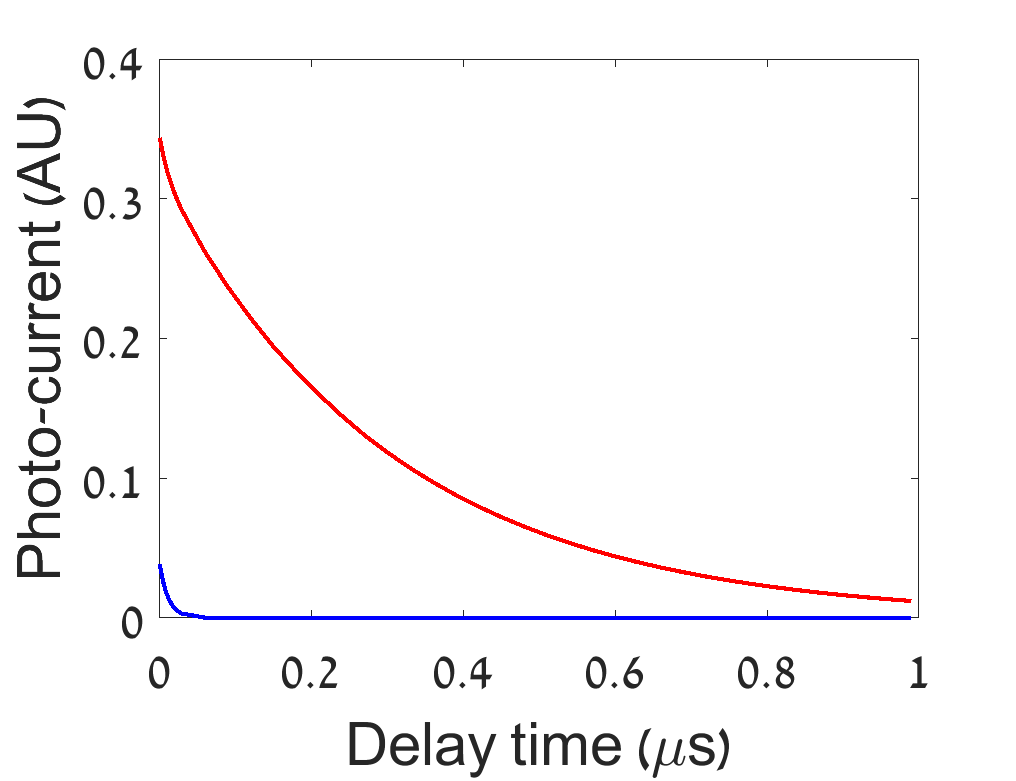}}
\subfigure[]
{\includegraphics[width = 0.45 \linewidth]{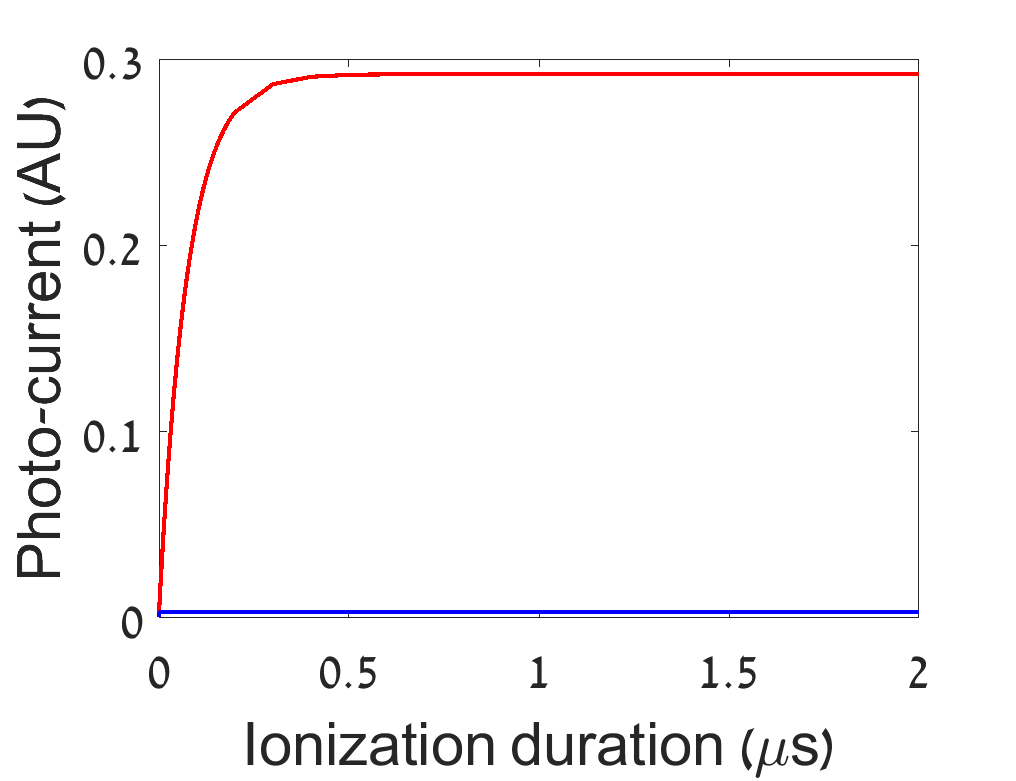}}

\caption{PC signal for various sequence parameters when 637 nm $ \leq  \lambda_{ion} \leq $ 674 nm. a. Even very low green population powers may generate high contrast in the signal. b. PC signal vs red excitation power. Higher power provides higher contrast, due to the delay between the green population pulse and the ionization pulse. c. NV$^-$ population vs singlet ionization cross section (in MHz/mW). Even very low cross section is expected to suffice, as almost no signal is expected under the threshold wavelength. d. PC signal vs the delay between the population and ionization pulses. The fast dynamics (short delays) stem from ionization from the triplet states. When longer delays are applied decreases with a time scale of the singlet lifetime. e. PC signal vs ionization pulse duration. The contrast builds as long as the singlet state is occupied, as no other ionization mechanism exists.} 

\label{fig:PC_Red}
\end{figure}

As before, we start with the longest wavelength range - 637 nm $ \leq  \lambda_{ion} \leq $ 674 nm. Fig. \ref{fig:PC_Red} depicts the expected behavior of the PC signal as a function of various parameters. The fact that the ionization laser does not excite the NV$^-$ triplet ground state or ionize P1 centers results in a rather 'clean' signal, as almost no PC is generated when the laser energy is lower than the ionization energy. Fig. \ref{fig:PC_Red}.a draws the PC signal as a function of green population power. The red curve trend changes according to the change in the singlet state population with green power, as high powers start to ionize the singlet state as well, resulting in lower population in this state. Increasing the ionization power, however, shows a monotonic behavior, as only the singlet ionization rate changes with this laser power while the blue curve is almost not affected (see Fig. \ref{fig:PC_Red}.b). Similarly to the fluorescence based measurement, the signal as a function of ionization cross section, depicted in Fig.  \ref{fig:PC_Red}.c, resembles the behavior as a function of green power, 
as although it causes higher ionization from the singlet during the ionization pulse, it reduces the population in the singlet prior to the ionization step due to green ionization during the population step. Fig. \ref{fig:PC_Red}.d shows the PC signal as a function of delay time between the population and ionization steps. For delays longer than few tens of ns, the red PC signal drops as the singlet lifetime, while for short times the signal decreases with the timescale of the triplet excited state lifetime for both curves. Increasing the ionization duration increases the PC until no population remains in the singlet state. However, other current inducing mechanisms, not included in this model, may provide an optimal duration for this step. 

\begin{figure}[tbh]
\subfigure[]
{\includegraphics[width = 0.45 \linewidth]{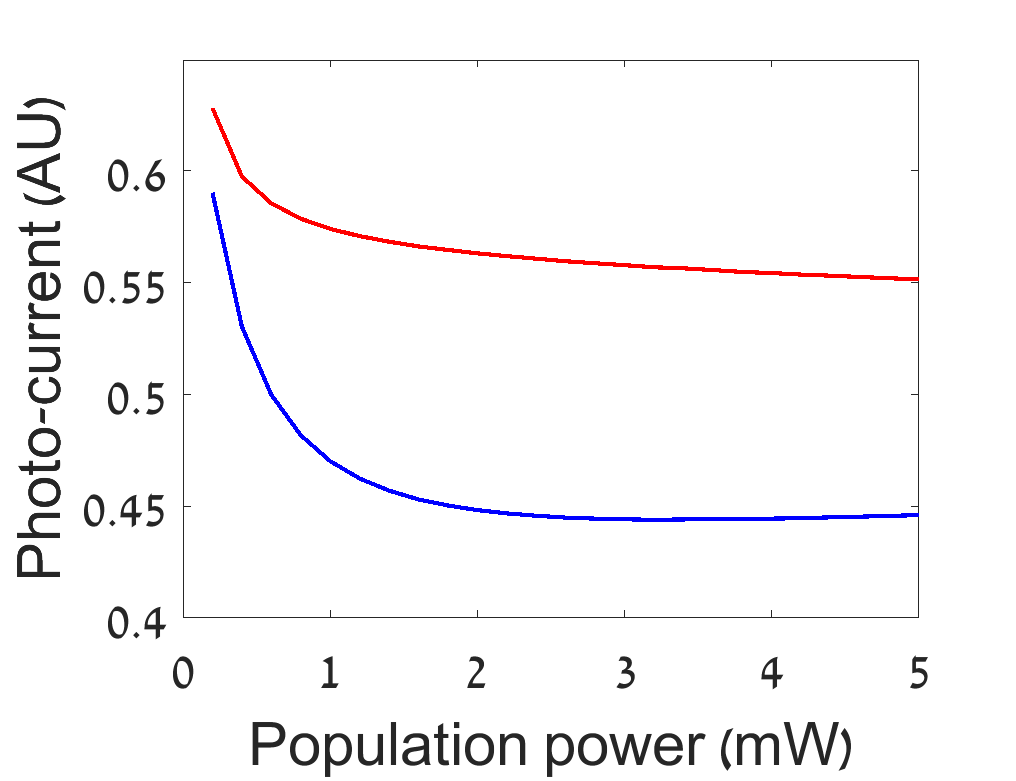}}
\subfigure[]
{\includegraphics[width = 0.45 \linewidth]{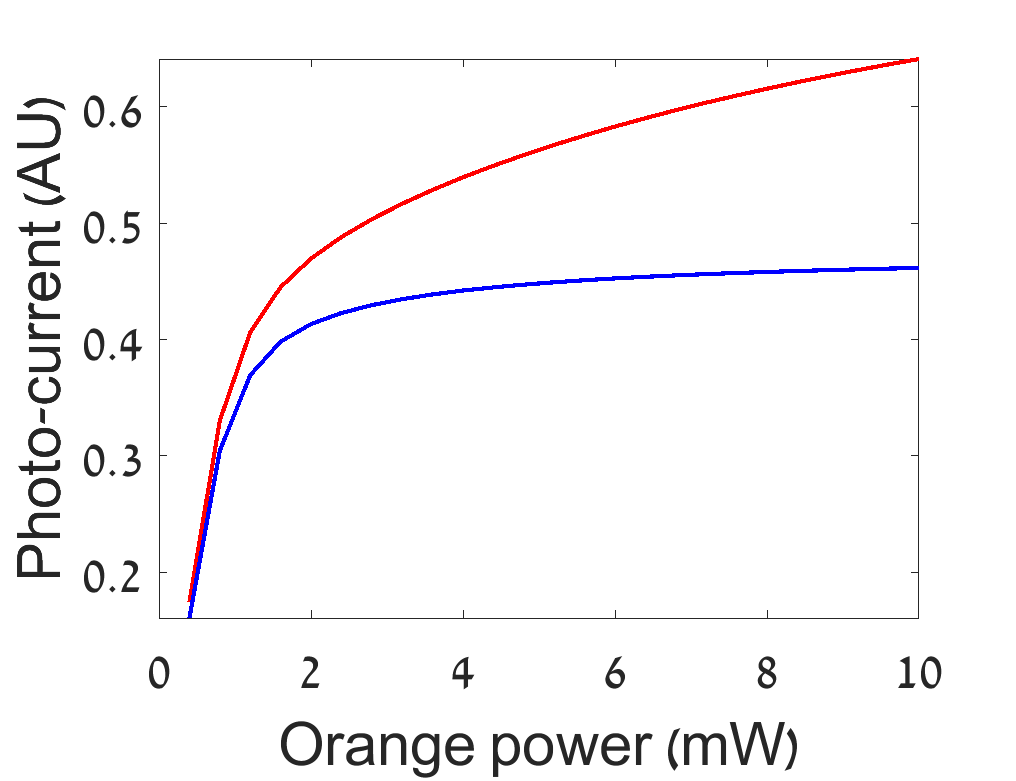}}
\subfigure[]
{\includegraphics[width = 0.45 \linewidth]{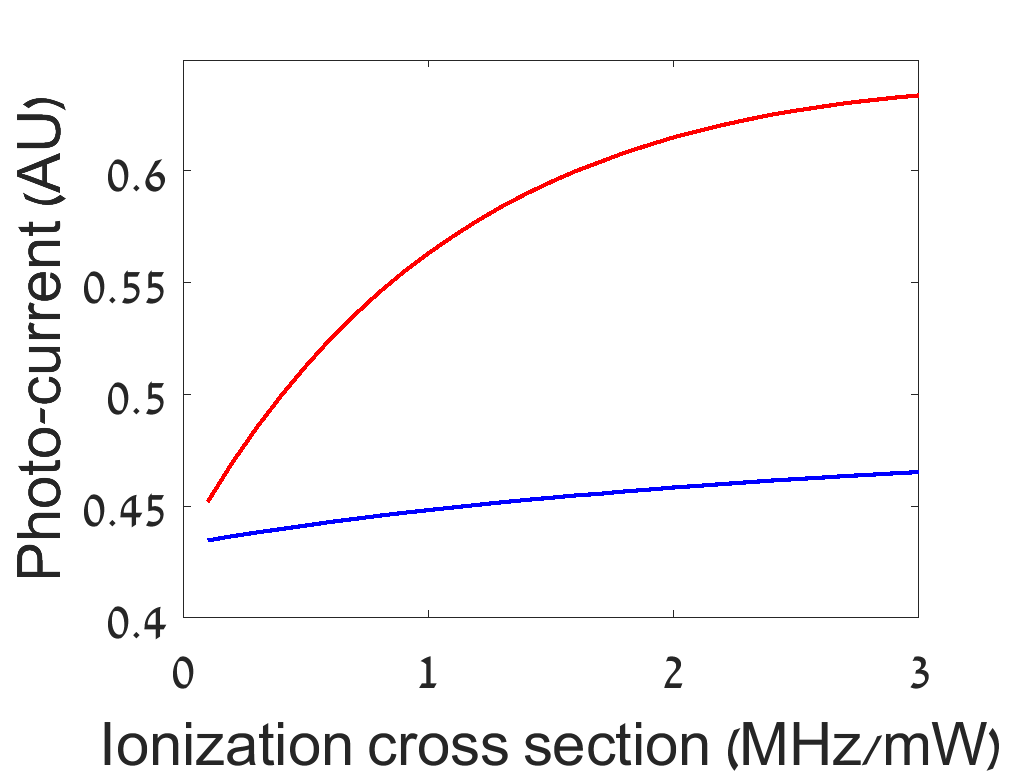}}
\subfigure[]
{\includegraphics[width = 0.45 \linewidth]{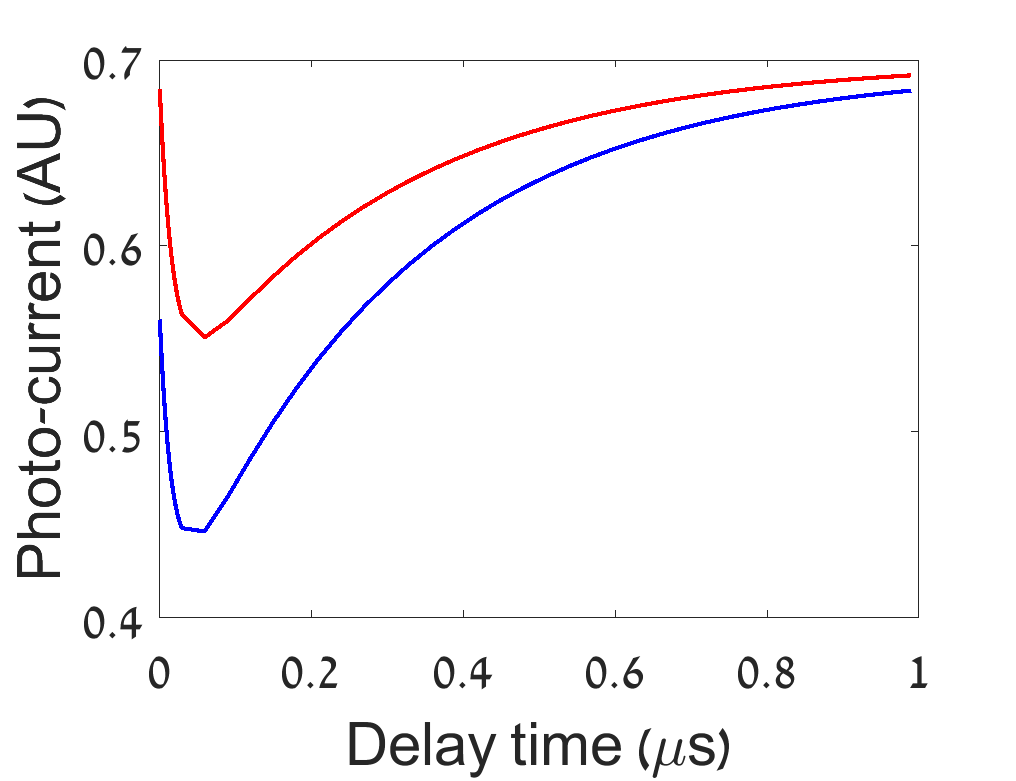}}
\subfigure[]
{\includegraphics[width = 0.45 \linewidth]{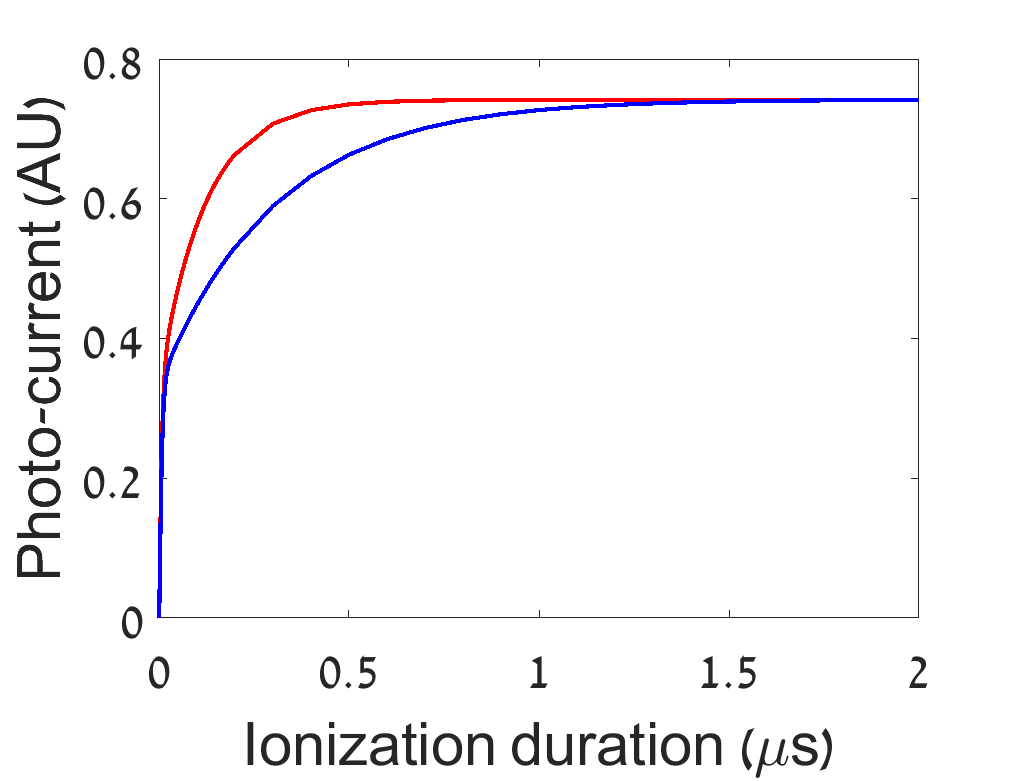}}

\caption{PC signal for various sequence parameters, where 637 nm $\leq \lambda_s \leq $ 575 nm. a. PC signal vs green power, populating the singlet level. powers $\leq$ 1 mW are preferred for significant contrast. b. PC signal vs orange excitation power. Higher power provides higher contrast, as the population in the triplet is expected to empty before the population in the singlet due to the assumption of low ionization cross section in the singlet. c. NV$^-$ population vs singlet ionization cross section (in MHz/mW). Ionization rate higher than 0.5 MHz/mW is expected to result in detectable contrast. d. PC signal vs the delay between the population and ionization pulses. The fast dynamics (short delays) stem from ionization from the triplet states. When longer delays are applied decreases with a time scale of the singlet lifetime. e. PC signal vs ionization pulse duration. The ionization duration must be similar to the singlet lifetime, otherwise no significant contrast is expected.}
\label{fig:PC_Orange}
\end{figure}

Moving to the second range, 575 nm $ \leq  \lambda_{ion} \leq $ 637 nm, we start seeing significant signal when no ionization from the singlet occurs as well, due to two-step ionization induced by the orange laser. Fig. \ref{fig:PC_Orange}.a shows a preference to higher laser powers with a weak optimum. In both curves most of the signal stems from the two-step ionization of the population in the triplet state, and the difference is proportional to the population in the singlet. At very high powers the singlet is less populated because of the high ionization rate, bringing the curves closer to each other. Fig. \ref{fig:PC_Orange}.b depicts the PC signal as a function of the orange power. While low powers produce similar signal for both scenarios, as the orange power increases the blue curve remains constant (since most of the triplet population has already been ionized), while the red curve continues to increase due to ionization from the singlet, creating the contrast between the two curves.
For measurements in this regime, one can use the blue curve's plateau as a baseline, to increase sensitivity for the ionization signal. As expected, higher cross section results in higher signal, and 0.5 MHz/mW should suffice for detectable contrast (Fig. \ref{fig:PC_Orange}.c). Both curves are affected by the cross section, as it changes the ionization during the population step (see Fig. \ref{fig:PC_Orange}.c). Fig. \ref{fig:PC_Orange}.d shows the PC signal as a function of delay time. Similar to Fig. \ref{fig:PC_Red}.d, the fast decrease in short times occurs due to the short lifetime of the triplet excited state. Later, the slow increase happens due to the singlet lifetime, since the simulations assume slower ionization from the singlet state than two-step ionization from the ground state. Lastly, Fig. \ref{fig:PC_Orange}.e depicts the PC signal as a function of ionization duration, showing a very similar behavior to Fig. \ref{fig:PL_Orange}.e, limiting our ionization duration for detectable contrast. However, this measurement can be substantially improved just by selecting the correct time to start the measurement. Both curves start with a fast increase in the PC signal, which stems almost entirely from ionization from the triplet excited state. Starting the measurement at the end of the fast step would eliminate most of the triplet ionization signal, thus enhance the contrast once ionization from the singlet occurs.

\begin{figure}[tbh]
\subfigure[]
{\includegraphics[width = 0.45 \linewidth]{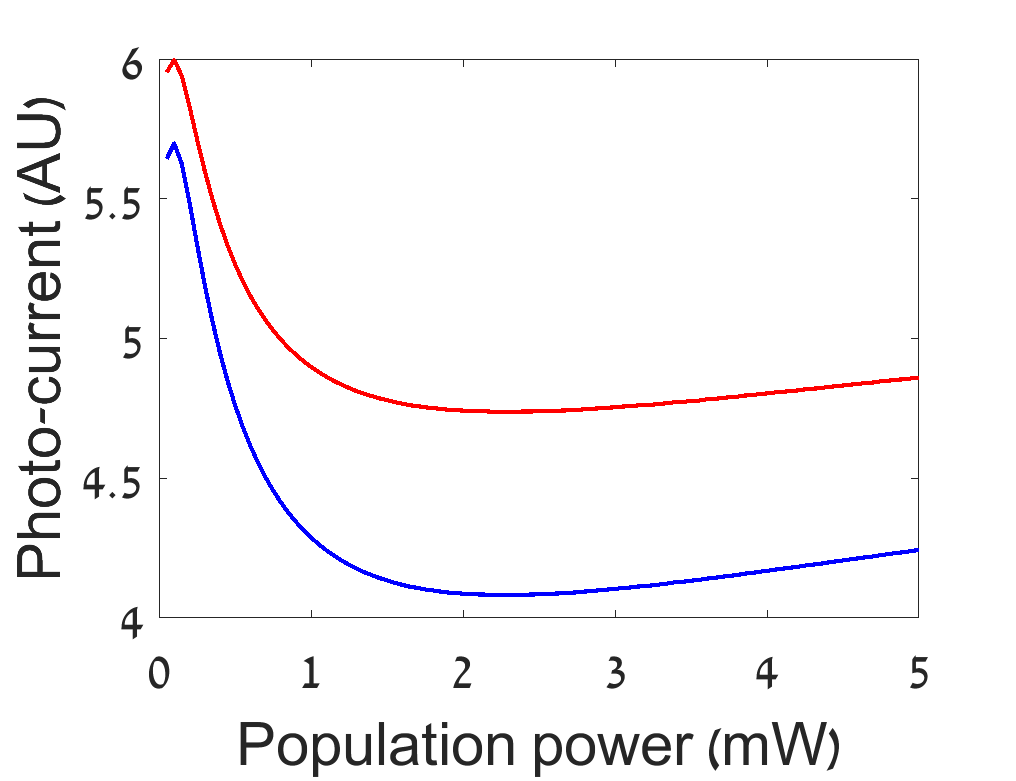}}
\subfigure[]
{\includegraphics[width = 0.45 \linewidth]{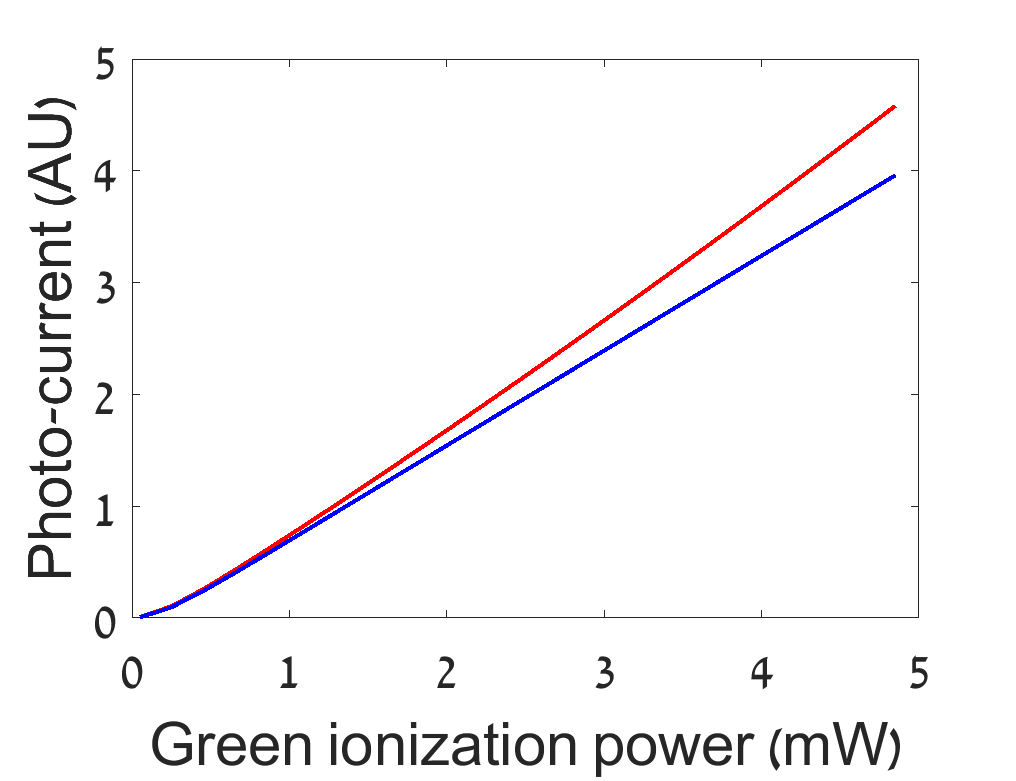}}
\subfigure[]
{\includegraphics[width = 0.45 \linewidth]{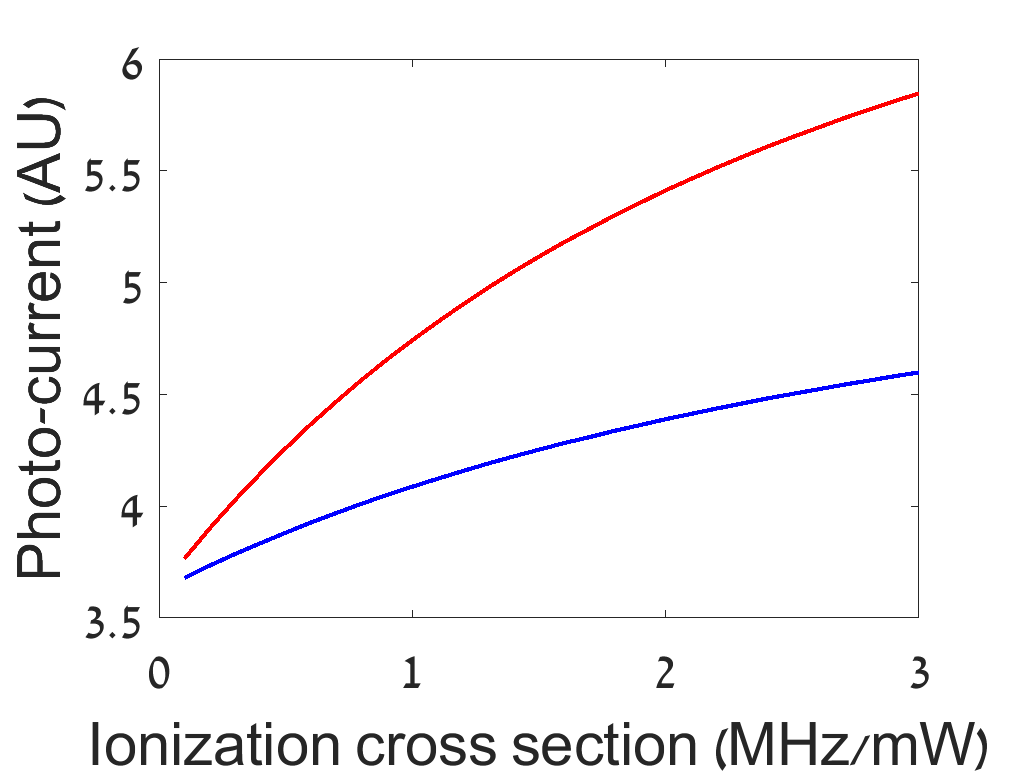}}
\subfigure[]
{\includegraphics[width = 0.45 \linewidth]{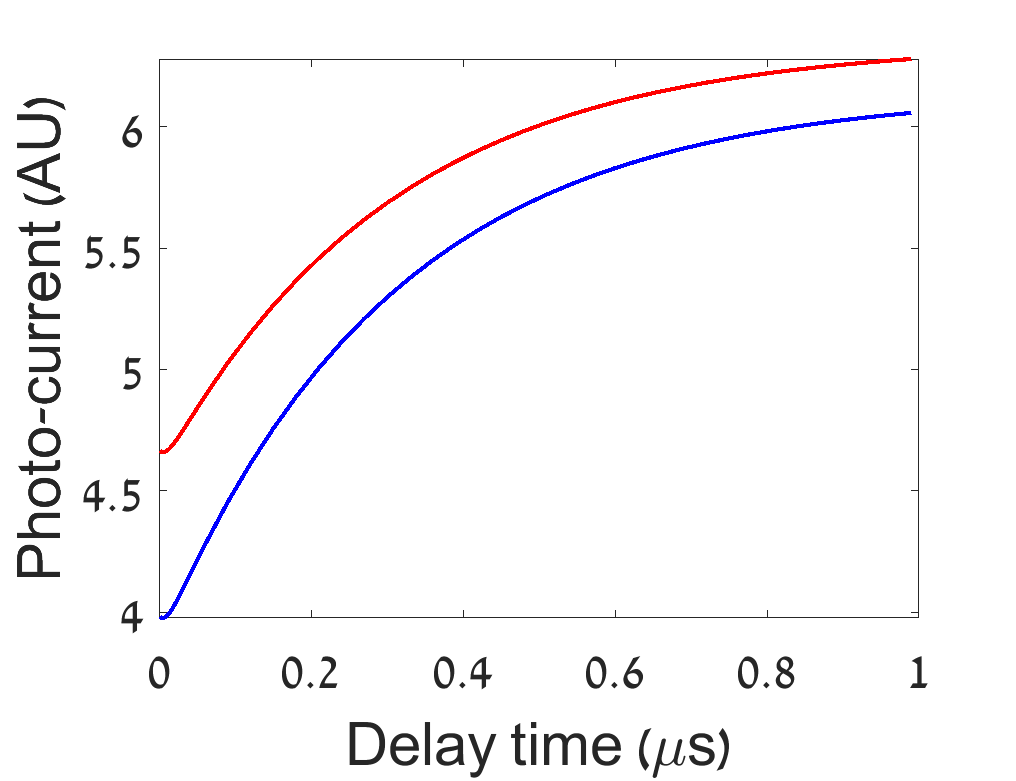}}
\subfigure[]
{\includegraphics[width = 0.45 \linewidth]{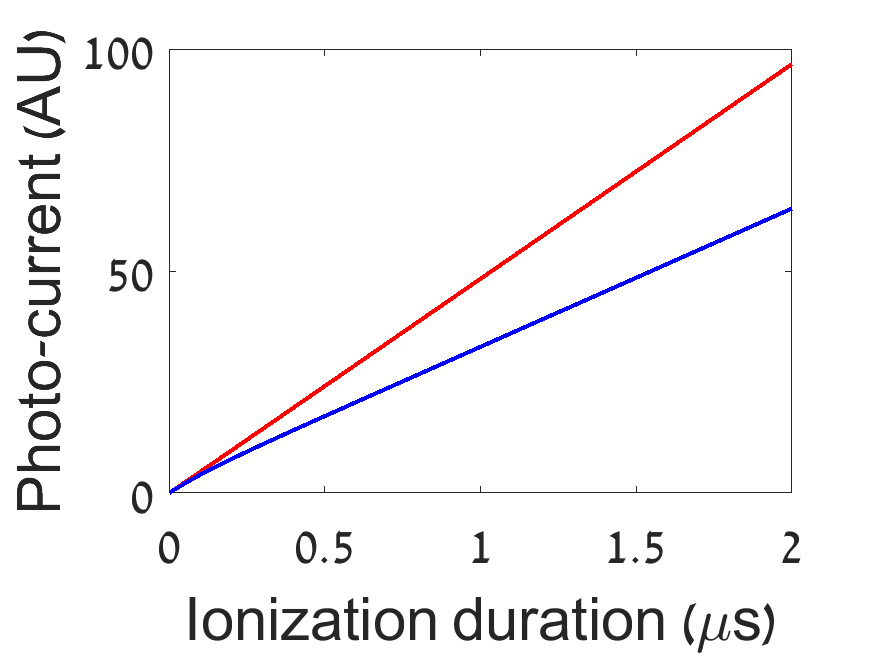}}

\caption{PC signal for various sequence parameters when 532 nm $  \leq \lambda_s \leq $ 575 nm. a. PC signal vs green population pulse power. A power $\leq$ 1 mW is preferred for significant contrast. b. PC signal vs green ionization power. Higher power provides higher contrast, as the singlet ionization cross section is expected to be smaller that the excited state cross section. c. PC signal vs singlet ionization cross section (in MHz/mW). Ionization rate higher than 0.5 MHz/mW is expected to result in detectable contrast. d. PC signal vs the delay between the population and ionization pulses. The contrast decreases with a time scale of the singlet lifetime. e. PC signal vs ionization pulse duration. The ionization duration is not bounded from above due to the fact that the population differences are not transient. The contrast builds up to 30\%.} 

\label{fig:PC_Green}
\end{figure}

The next range, 532 nm $ \leq  \lambda_{ion} \leq $ 575 nm, shows a simpler behavior compared to the previous range. Fig. \ref{fig:PC_Green}.a resembles Fig. \ref{fig:PC_Orange}.a, as the main source of contrast in this figure comes from the population of the singlet state, while at some point the population laser start ionizing the NV from the singlet. However, Fig. \ref{fig:PC_Green}.b shows much less contrast as a function of ionization power compared to Fig. \ref{fig:PC_Orange}.b, because in this case the wavelength is below 575 nm, enabling two step recombination process. This causes multiple ionization of the same NV, and due to the assumption that the two-step ionization process from the ground state would be faster than the ionization from the singlet, this results in a lower contrast of the PC signal. Similarly to the previous figures, a cross section corresponding to a rate higher than 0.5 MHz/mW should provide a detectable contrast when reaching the threshold ionization wavelength, while the delay time between the population and ionization pulses should be below 200 ns for optimal results, as depicted in Figs. \ref{fig:PC_Green}.c and \ref{fig:PC_Green}.d respectively.
Lastly, Fig. \ref{fig:PC_Green}.e shows higher signal as a function of ionization duration, while the contrast reaches a plateau. This is how the steady state charge-state, described in \ref{fig:PL_Green}.e, is manifested in the PC signal. 

\begin{figure}[tbh]
\subfigure[]
{\includegraphics[width = 0.45 \linewidth]{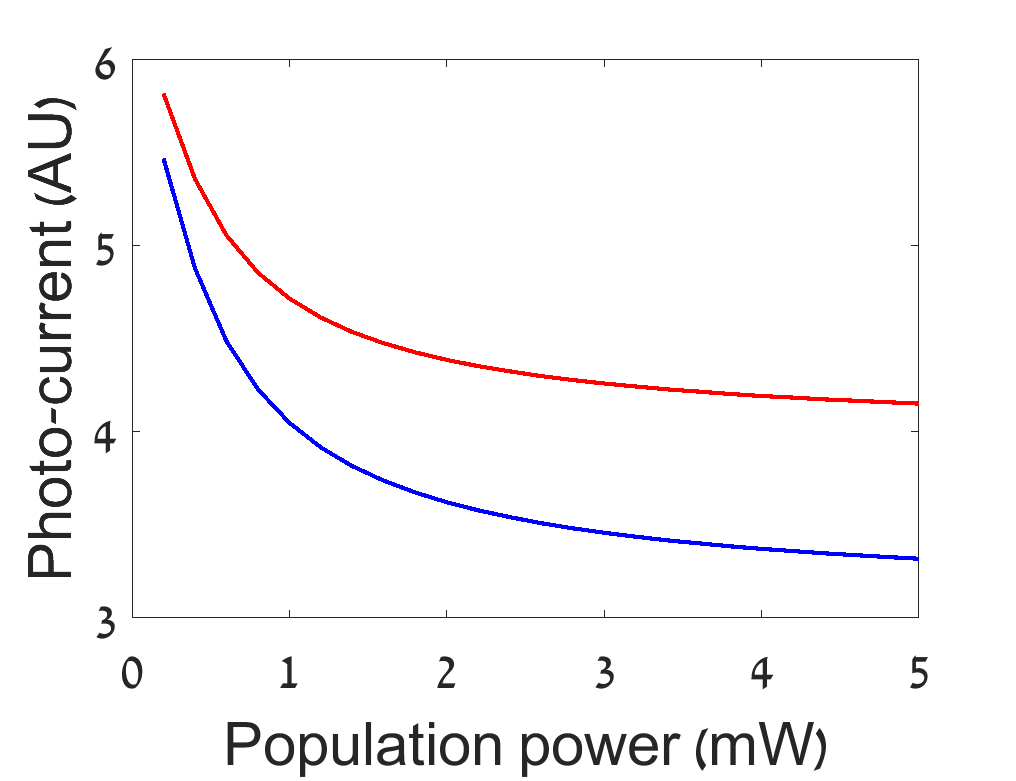}}
\subfigure[]
{\includegraphics[width = 0.45 \linewidth]{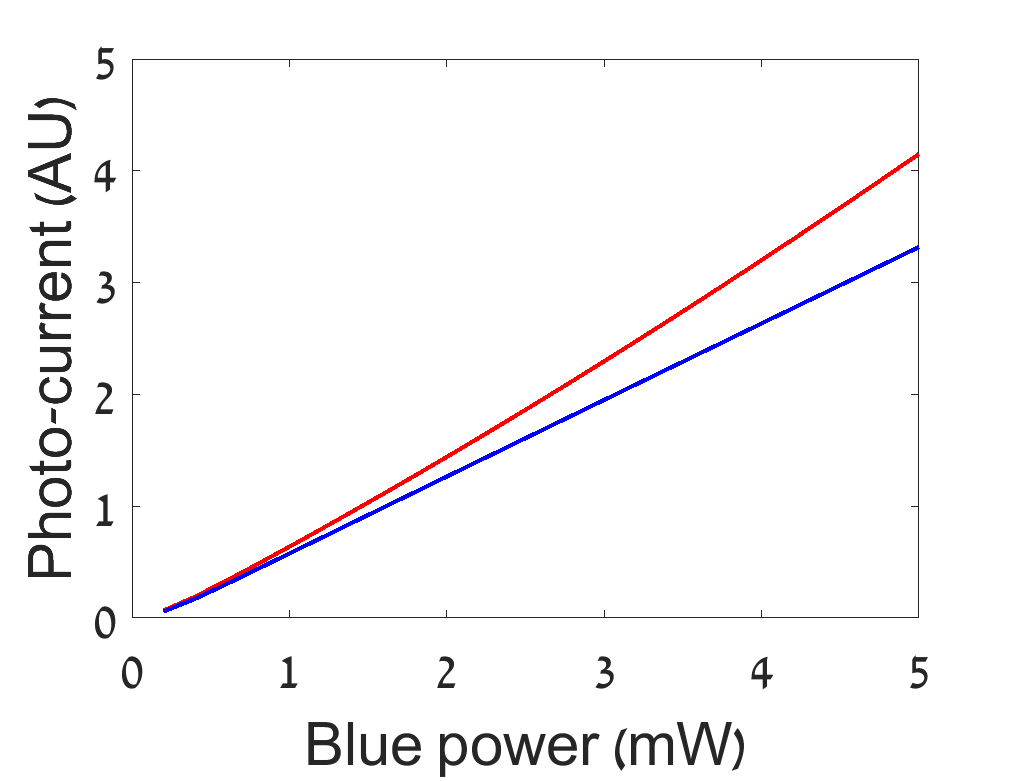}}
\subfigure[]
{\includegraphics[width = 0.45 \linewidth]{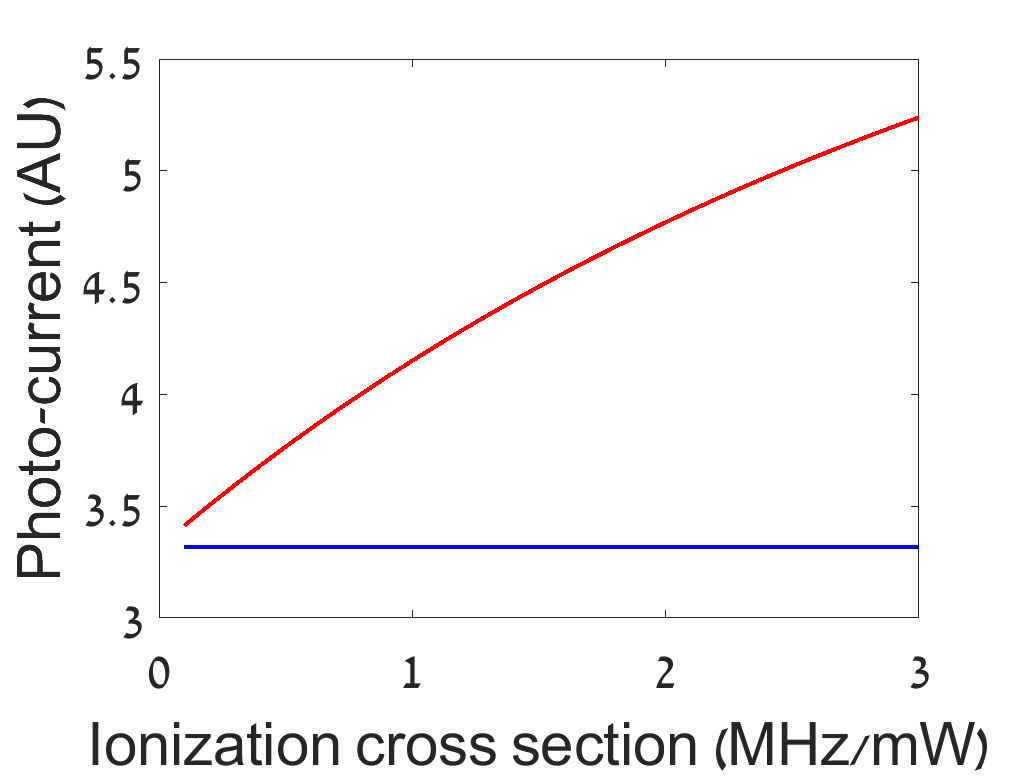}}
\subfigure[]
{\includegraphics[width = 0.45 \linewidth]{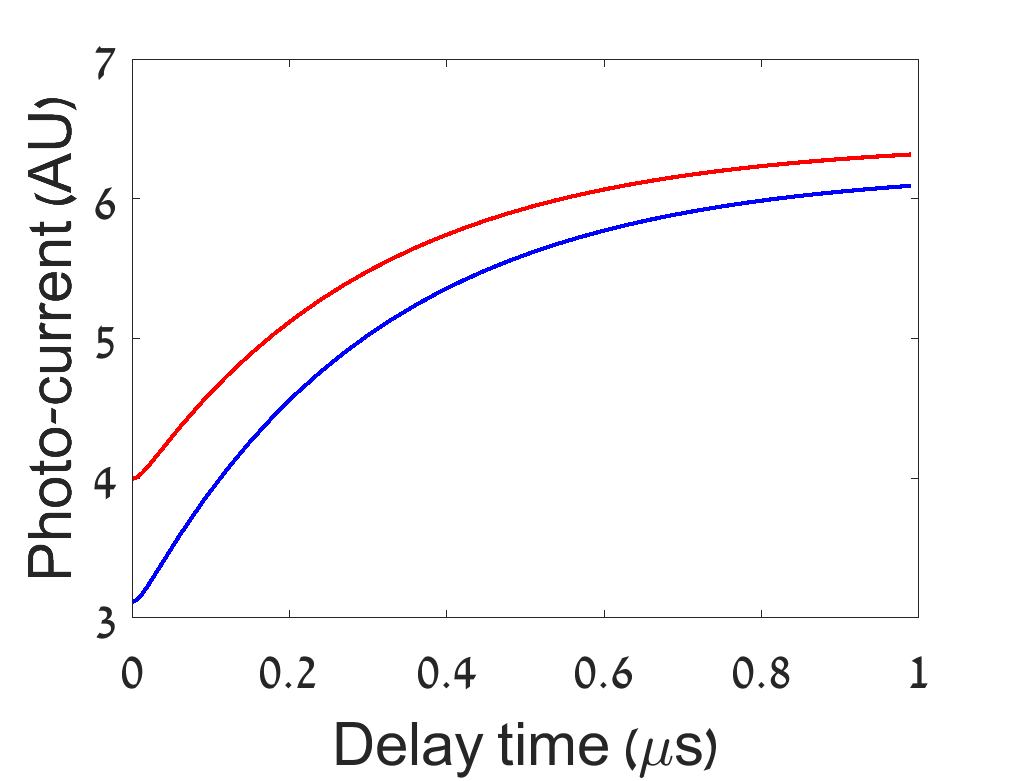}}
\subfigure[]
{\includegraphics[width = 0.45 \linewidth]{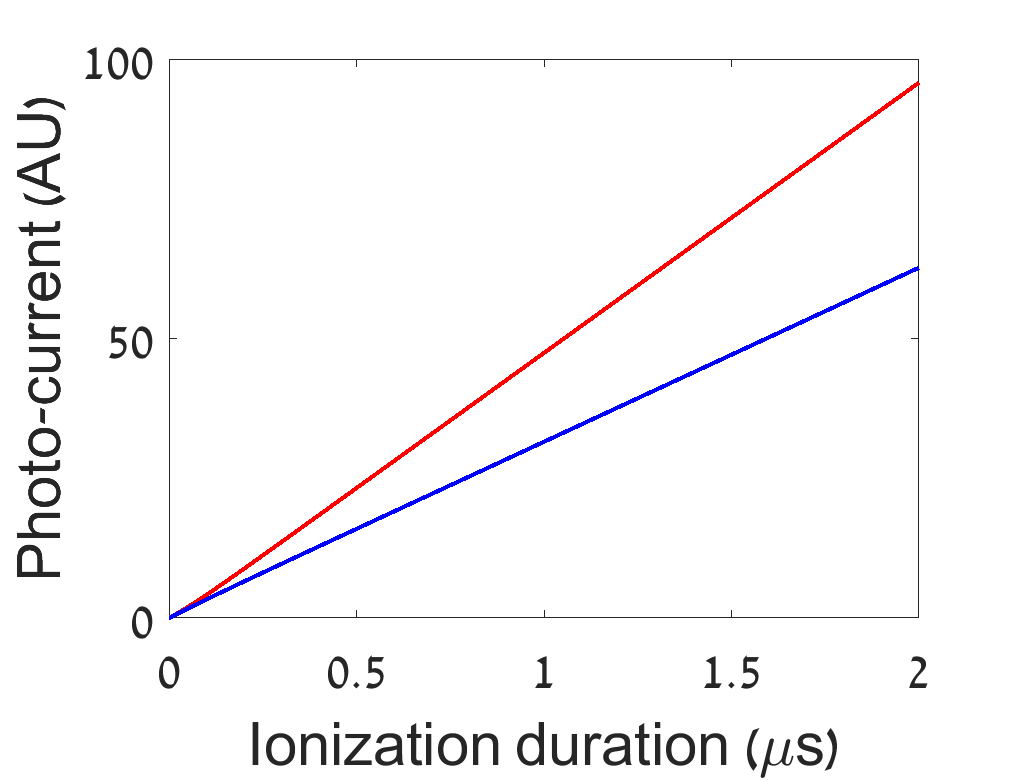}}

\caption{PC signal for various sequence parameters when 477 nm $ \leq \lambda_s \leq $ 532 nm, showing almost identical behavior as in Fig. \ref{fig:PC_Green}. a. PC signal vs green population pulse power. A power $\leq$ 1 mW is preferred for significant contrast. b. PC signal vs the green ionization power. Higher power provides higher contrast, as the singlet ionization cross section is expected to be smaller that the excited state cross section. c. PC signal vs singlet ionization cross section (in MHz/mW). Ionization rate higher than 0.5 MHz/mW is expected to result in detectable contrast. d. PC signal vs the delay between the population and ionization pulses. The fast dynamics (short delays) stem from ionization from the triplet states. The contrast decreases with a time scale of the singlet lifetime. e. PC signal vs ionization pulse duration. The ionization duration is not bounded from above due to the fact that the populations differences are not transient.} 

\label{fig:PC_Blue}
\end{figure}

Like the fluorescence based measurement, the last range - 477 nm $ \leq  \lambda_{ion} \leq $ 532 nm presents similar behavior as the previous range (see Fig. \ref{fig:PC_Blue}). 

\section{discussion}

Comparing the two detection methods, there is no clear advantage for one over the other for wavelengths below 637 nm. For wavelengths longer than 637 nm the PC approach is expected to produce significantly higher contrast since ionization from the excited state is expected to be eliminated. In addition, multiple cycles of population + ionization steps are likely to further enhance the signal without adversely affecting the contrast, as long as the population laser powers are carefully controlled. However, as the ionization energy increases, it induces additional ionization processes of other defects in the diamond lattice, thus masking the desired signal from the NV. 

Fluorescence-based measurements can be done by either reading the NV$^-$ signal or the NV$^0$. However, most of the cases covered in this work have shown limited contrast in the NV$^-$ population due to expected spin mixing. Additional spin mixing generated by the relatively high laser powers used in the sequences might eliminate the remaining contrast below detection threshold. Assuming high NV density, accurate NV$^0$ population measurements may result in higher SNR compared to collecting NV$^-$ fluorescence, despite the weaker signal, especially when working in the blue and green regimes. In these regimes, the NV strongly prefers the negative charge state, thus the 'jump' in the NV$^0$ population once reaching the threshold wavelength may provide substantially higher contrast, overcoming the reduced number of photons. 

The simulations assume a conservative approach regarding the lasers used during the experiment. Specifically, the dynamics presented consider CW lasers and a modulation component. However, using a pulsed laser (for the ionization step) holds the potential to boost the SNR of the measurements dramatically, if the pulse duration is short enough. As long as the laser is non-resonant with the $^3A \rightarrow ^3E$ transition, pulses shorter than $\sim$ 5 ps are not expected to cause substantial ionization from the $^3E$ state into the conduction band \cite{huxter_vibrational_2013}. Thus, most of the ionization will originate from NVs in the singlet state. In this case, the blue curves in the PC section become negligible at all wavelength ranges (as in Fig. \ref{fig:PC_Red}), while the red curves in the fluorescence section drop (as in Fig. \ref{fig:PL_Red}). This is, of course, as long as the repetition time is $\simleq$ 30 ns, such that the $^3E$ state decays almost fully between pulses. 

\section{Acknowledgements}

This work has been supported by the Ministry of Science and Technology, Israel, and the European Union’s Horizon 2020 research and innovation program under grant agreements No. 714005 (ERC StG Q-DIM-SIM), No. 820374 (MetaboliQs), and No. 828946 (PATHOS).



\bibliography{main.bbl}

\end{document}